\documentclass[preprint,aps,prd,showpacs,nofootinbib,superscriptaddress,12pt]{revtex4}%
\usepackage{graphicx}
\usepackage{amsmath}
\usepackage{amssymb}
\usepackage{bm}
\usepackage{epsfig}
\usepackage{epstopdf}
\begin{document}
\title{\mbox{}\\[10pt]
Fragmentation contributions to hadroproduction of 
\\prompt $\bm{J/\psi}$, $\bm{\chi_{cJ}}$, and $\bm{\psi(2S)}$ states
}
\author{Geoffrey~T.~Bodwin}
\affiliation{High Energy Physics Division, Argonne National Laboratory,\\
9700 South Cass Avenue, Argonne, Illinois 60439, USA}
\author{Kuang-Ta~Chao}
\affiliation{School of Physics and State Key Laboratory of Nuclear Physics and
Technology, Peking University, Beijing 100871, China}
\affiliation{Center for High Energy physics,
Peking University, Beijing 100871, China}
\affiliation{Collaborative Innovation Center of Quantum Matter,
Beijing 100871, China}
\author{Hee~Sok~Chung}
\affiliation{High Energy Physics Division, Argonne National Laboratory,\\
9700 South Cass Avenue, Argonne, Illinois 60439, USA}
\author{U-Rae~Kim}
\affiliation{Department of Physics, Korea University, Seoul 136-701, Korea}
\affiliation{Korea Institute for Advanced Study, Seoul 130-722, Korea}
\author{Jungil~Lee}
\affiliation{Department of Physics, Korea University, Seoul 136-701, Korea}
\author{Yan-Qing~Ma}
\affiliation{School of Physics and State Key Laboratory of Nuclear Physics and
Technology, Peking University, Beijing 100871, China}
\affiliation{Collaborative Innovation Center of Quantum Matter,
Beijing 100871, China}

\date{\today}
\begin{abstract}

We compute fragmentation corrections to hadroproduction of the
quarkonium states $J/\psi$, $\chi_{cJ}$, and $\psi(2S)$ at leading power
in $m_c^2/p_T^2$, where $m_c$ is the charm-quark mass and $p_T$ is the
quarkonium transverse momentum. The computation is carried out in the
framework of nonrelativistic QCD. We include corrections to the
parton-production cross sections through next-to-leading order in the
strong coupling $\alpha_s$ and corrections to the fragmentation
functions through second order in $\alpha_s$. We also sum leading
logarithms of $p_T^2/m_c^2$ to all orders in perturbation theory.  We
find that, when we combine these leading-power fragmentation
corrections with fixed-order calculations through next-to-leading order
in $\alpha_s$, we are able to obtain good fits for $p_T\geq 10$~GeV to
hadroproduction cross sections that were measured at the Tevatron and
the LHC. Using values for the nonperturbative long-distance matrix
elements that we extract from the cross-section fits, we make
predictions for the polarizations of the quarkonium states. We obtain
good agreement with measurements of the polarizations, with the
exception of the CDF Run~II measurement of the prompt $J/\psi$
polarization, for which the agreement is only fair. In the predictions
for the prompt-$J/\psi$ cross sections and polarizations, we take
into account feeddown from the $\chi_{cJ}$ and $\psi(2S)$ states.

\end{abstract}
\maketitle

\section{Introduction}

In recent years, corrections to inclusive quarkonium production cross
sections and polarizations
through next-to-leading order (NLO) in the strong coupling $\alpha_s$
have been computed for both hadroproduction \cite{Ma:2010jj,Butenschoen:2010rq,Ma:2010yw,Chao:2012iv,
Butenschoen:2012px,Gong:2012ug} and
photoproduction \cite{Butenschoen:2009zy,Butenschoen:2011ks,Butenschoen:2011yh}.
These computations have been carried out in the context of
the nonrelativistic QCD (NRQCD) factorization
conjecture~\cite{Bodwin:1994jh}, which states that
the inclusive production cross section to produce a quarkonium $H$ in a
collision of particles $A$ and $B$ can be written as
\begin{equation}
d \sigma_{A+B \to H + X}= \sum_n d \sigma_{A+B \to Q \bar Q(n) + X}
\langle  {\cal O}^H(n) \rangle.
\label{NRQCD-fact}
\end{equation}
Here, the $d \sigma_{A+B \to Q \bar Q(n)+X}$ are the short-distance
coefficients (SDCs), which can be computed in perturbation theory and
which correspond to the production of
a heavy quark-antiquark pair $Q \bar Q(n)$ in a specific
color and angular-momentum state $n$.  The $\langle {\cal O}^H(n)
\rangle$ are NRQCD long-distance matrix elements
(LDMEs), which parametrize the nonperturbative part of the production
process.

Because the LDMEs have a known scaling with $v$, the heavy-quark
velocity in the quarkonium rest frame \cite{Bodwin:1994jh}, the sum in
Eq.~(\ref{NRQCD-fact}) can be regarded as an expansion in the small
parameter $v$. ($v^2\approx 0.3$ for the $J/\psi$.) In present-day
phenomenology, the sum in Eq.~(\ref{NRQCD-fact}) is truncated at
relative order $v^4$. For $H = J/\psi$ or $H=\psi(2S)$, the truncated
sum involves four LDMEs: $\langle{\cal
O}^{\psi}({}^3S_1^{[1]})\rangle$, $\langle{\cal
O}^{\psi}({}^3S_1^{[8]})\rangle$, $\langle{\cal
O}^{\psi}({}^1S_0^{[8]})\rangle$, and $\langle{\cal
O}^{\psi}({}^3P_J^{[8]})\rangle$, where the expressions in parentheses
give the color state of the $Q\bar Q$ pair (singlet or octet) and spin
and orbital angular momentum in spectroscopic notation. Here, $\psi$ stands
for $J/\psi$ or $\psi(2S)$. For
$H=\chi_{cJ}$, the truncated sum involves two LDMEs: $\langle{\cal
O}^{\chi_{c0}}({}^3P_0^{[1]})\rangle$ and $\langle{\cal
O}^{\chi_{c0}}({}^3S_1^{[8]})\rangle$, where the LDMEs for the $\chi_{c1}$
and $\chi_{c2}$ states can be related to the LDMEs for the $\chi_{c0}$ state
by making use of the heavy-quark spin symmetry \cite{Bodwin:1994jh},
which is valid up to corrections of relative order $v^2$.

Since the color-singlet LDME for quarkonium production $\langle {\cal
O}^{\psi} ({}^3S_1^{[1]}) \rangle$ is related to the color-singlet LDME
for quarkonium decay, it can be determined in lattice QCD, from
potential models, or from the $\psi$ decay rates into lepton pairs. On
the other hand, it is not known how to compute the color-octet
production LDMEs from first principles, and they are usually fixed
by comparisons of NRQCD factorization predictions with measured cross
sections.

Even at the level of NLO accuracy in the theoretical predictions, it is
not possible to achieve a fully consistent description of the existing
$J/\psi$ production data within the NRQCD framework. For example, one
can fit the hadroproduction cross-section data
\cite{Acosta:2004yw,Chatrchyan:2011kc} and polarization data
\cite{Affolder:2000nn,Abulencia:2007us,Chatrchyan:2013cla}
simultaneously \cite{Chao:2012iv}, but the LDMEs that are obtained yield
a prediction for the photoproduction cross section that is larger than
the HERA data from the H1 Collaboration
\cite{Adloff:2002ex,Aaron:2010gz} by factors of 4--8 at the highest
value of $p_T$ at which the cross section has been measured
\cite{Butenschoen:2012qr}. On the other hand, one can fit the
predictions for the hadroproduction and photoproduction cross sections
to the experimental data \cite{Butenschoen:2012px}, but the LDMEs that
are obtained lead to predictions of large transverse polarization in
hadroproduction at large $p_T$, in disagreement with the experimental
data \cite{Butenschoen:2012px}. In addition, it was found in
Ref.~\cite{Butenschoen:2014dra} that the $\eta_c$ production data that
were measured by the LHCb Collaboration \cite{Aaij:2014bga} are
incompatible with the LDMEs that were extracted in
Ref.~\cite{Butenschoen:2012px} from hadroproduction and photoproduction
cross-section data.  Although one can describe the $\eta_c$ production
data by using the LDMEs that were extracted in Ref.~\cite{Ma:2010yw},
there is a very large cancellation between the contributions from the
${}^3S_1^{[8]}$ and ${}^3P_J^{[8]}$ channels
\cite{Han:2014jya,Zhang:2014ybe}, and, hence, the remainder may be
strongly dependent on uncertainties from uncalculated higher-order
contributions.

These difficulties provide motivation for calculations of quarkonium
production cross sections beyond NLO accuracy in $\alpha_s$. An approach
that simplifies computations beyond NLO in $\alpha_s$ is to compute
rates at leading power (LP) or next-to-leading power (NLP) in
$m_c^2/p_T^2$, where $m_c$ is the charm-quark mass and $p_T$ is the
quarkonium transverse momentum. LP contributions can be factorized into
semi-inclusive partonic cross sections to produce a specific single
parton convolved with one-parton fragmentation functions (FFs)
\cite{Collins:1981uw}. NLP contributions can be factorized into
semi-inclusive partonic cross sections to produce two specific partons
convolved with two-parton FFs \cite{Kang:2011mg}.
Calculations of these fragmentation contributions, at any given
order in $\alpha_s$, are much simpler than a full fixed-order
calculation. Furthermore, the LP- and NLP-factorization frameworks are
natural ones within which to resum large logarithms of $p_T^2/m_c^2$. Of
course, because the LP and NLP contributions represent
the leading and first subleading terms in an expansion in powers of
$m_c^2/p_T^2$, one would not expect them to be valid unless $p_T$ is
significantly greater than $m_c$.

In Ref.~\cite{Bodwin:2014gia} it was found that LP contributions beyond
NLO in $\alpha_s$ are important in $J/\psi$ hadroproduction. With the
inclusion of these contributions, the LDMEs that are extracted from the
prompt hadroproduction cross sections alone yield predictions for the
$J/\psi$ polarization at large $p_T$ that are near zero and are in
agreement with the experimental data \cite{Bodwin:2014gia}. One
deficiency in the analysis of Ref.~\cite{Bodwin:2014gia} is that it does
not take into account the effects of feeddown from the $\chi_{cJ}$ and
$\psi(2S)$ states to the $J/\psi$.

In this paper, we remedy that deficiency and extend the application of
the LP-factorization approach by computing LP-fragmentation
contributions to direct $J/\psi$, $\chi_{cJ}$, and $\psi(2S)$
production. We extract LDMEs by fitting to the Tevatron and LHC
production cross sections, and we use those LDMEs to predict the
$J/\psi$, $\chi_{cJ}$, and $\psi(2S)$ polarizations. Our predictions for
the prompt $J/\psi$ and $\psi(2S)$ polarizations agree well with the
existing high-$p_T$ LHC data, but the prompt $J/\psi$  polarization is
in only fair agreement with the high-$p_T$ Tevatron Run~II data. Our
predictions for the $\chi_{cJ}$ polarizations will be tested soon at the
LHC. While the results in this paper do not resolve the discrepancies
between the NRQCD predictions and the $J/\psi$ photoproduction and
$\eta_c$ hadroproduction data, they do provide a consistent description
of the existing spin-triplet charmonium hadroproduction data at high
$p_T$.

The remainder of this paper is organized as follows. In
Sec.~\ref{sec:LP-corrections}, we discuss the form of the LP corrections
that we compute. Section~\ref{sec:LP-SDCs} contains the details of the
calculation of the LP SDCs. We combine the LP and NLO results for the
SDCs in Sec.~\ref{sec:LP+NLO}. In Sec.~\ref{sec:LDMEs}, we fit our
predictions for the hadroproduction cross sections to the data,
obtaining values for the LDMEs. We use these values for the LDMEs
to make predictions for cross-section ratios and polarizations in
Sec.~\ref{sec:predictions}. Finally, in Sec.~\ref{sec:summary}, we
summarize and discuss our results.

\section{Corrections to Quarkonium Production at Leading Power in
$\bm{p_T}$\label{sec:LP-corrections}}

The contribution of leading power in $p_T$ to a quarkonium production
cross section is given by the LP-factorization formula
\cite{Collins:1981uw}
\begin{equation}
d \sigma^{\rm LP}_{A+B \to Q \bar Q(n) + X}(p)
=
\int_0^1 dz
\sum_i d\hat\sigma_{A+B\to i+X}(p_i=p/z, \mu_f)
D_{i\to Q \bar Q(n)} (z, \mu_f).
\label{sigma-LP-fact}
\end{equation}
Here, $d\hat\sigma_{A+B\to i+X}$ is the
semi-inclusive parton-production cross section (PPCS) for hadrons $A$
and $B$ to produce parton $i$, and $D_{i\to Q \bar Q(n)}$ is
the FF for parton $i$ to fragment into the
$Q\bar Q$ pair with quantum numbers $n$. $p$ is the momentum of the $Q\bar Q$
pair, which is taken to be lightlike by neglecting the heavy-quark
mass, and $p_i$ is the momentum of parton $i$, which is taken to be
lightlike by neglecting the parton mass. $\mu_f$ is the factorization scale.

As we will describe in more detail in Sec.~\ref{sec:LP-SDCs},
the PPCSs and the FFs have been calculated to
order $\alpha_s^3$ and $\alpha_s^2$, respectively. Hence, we write them
as
\begin{subequations}
\begin{eqnarray}
d\hat\sigma_{A+B\to i+X}
&=&
\alpha_s^2 d\hat\sigma_{A+B\to i+X}^{(2)}
+
\alpha_s^3 d\hat\sigma_{A+B\to i+X}^{(3)}
+ O(\alpha_s^4),
\\
D_{i\to Q \bar Q(n)}
&=&
\alpha_s D_{i\to Q \bar Q(n)}^{(1)} +
\alpha_s^2 D_{i\to Q \bar Q(n)}^{(2)} +
O (\alpha_s^3).
\end{eqnarray}
\end{subequations}

As we have already mentioned, the SDCs for both unpolarized and
polarized quarkonium production have been computed through NLO in
$\alpha_s$, which is order $\alpha_s^4$. In this paper, we extend these
order-$\alpha_s^4$ calculations by combining existing calculations of the
PPCSs through order $\alpha_s^3$ and existing calculations of the FFs
through order $\alpha_s^2$ to obtain a partial calculation of the
order-$\alpha_s^5$ (NNLO) contributions to the LP SDCs. Furthermore, we
calculate corrections to the LP SDCs involving leading logarithms of
$p_T^2/m_c^2$ to all orders in $\alpha_s$ by solving the
Dokshitzer-Gribov-Lipatov-Altarelli-Parisi (DGLAP) evolution
equation~\cite{Gribov:1972ri, Lipatov:1974qm, Dokshitzer:1977sg,
Altarelli:1977zs}. Because this calculation of the LP SDCs accounts
only partially for corrections of order $\alpha_s^5$, we expect
uncertainties from uncalculated corrections to be of order $\alpha_s^5$.
However, these uncalculated corrections will not contain any
enhancements from leading logarithms of $p_T^2/m_c^2$.

Part of the LP-fragmentation contribution through order $\alpha_s^4$ is
already included in the NLO SDCs, namely,
\begin{eqnarray}
&&d \sigma^{\rm LP}_{\rm NLO}(p)
=
\int_0^1 dz \sum_i \alpha_s^3  d\hat\sigma_{A+B\to i+X}^{(2)}(p_i=p/z, \mu_f)
D_{i\to Q \bar Q(n)}^{(1)} (z, \mu_f)
\nonumber \\
&&
+\int_0^1 dz \sum_i \alpha_s^4
\left[ d\hat\sigma_{A+B\to i+X}^{(2)}(p_i=p/z, \mu_f)
D_{i\to Q \bar Q(n)}^{(2)} (z, \mu_f)
\right.
\nonumber \\
&&
\left. \hspace{18ex}
+
d\hat\sigma_{A+B\to i+X}^{(3)}(p_i=p/z, \mu_f)
D_{i\to Q \bar Q(n)}^{(1)} (z, \mu_f) \right].
\label{LP-NLO}
\end{eqnarray}
Hence, when we combine the SDCs through NLO in $\alpha_s$ and the
LP-fragmentation contributions, we must subtract the contributions in
Eq.~(\ref{LP-NLO}) in order to avoid double counting. Following
Ref.~\cite{Bodwin:2014gia}, we compute
\begin{equation}
\frac{d \sigma^{\rm LP+NLO}}{dp_T}
=
\frac{d \sigma^{\rm LP}}{dp_T}
- \frac{d \sigma^{\rm LP}_{\rm NLO}}{dp_T}
+ \frac{d \sigma_{\rm NLO}}{dp_T},
\label{LP+NLO}
\end{equation}
where $d \sigma_{\rm NLO}/dp_T$ is the SDC through NLO in $\alpha_s$.
The expression (\ref{LP+NLO}) takes into account, without double
counting, the complete calculations through NLO in $\alpha_s$ and also
the additional LP corrections beyond NLO that we have mentioned.

\section{Computation of the LP short-distance coefficients
\label{sec:LP-SDCs}}

In this section we describe the details of the computation of the
PPCSs and FFs
that enter into the LP short-distance coefficients in the LP
factorization formula (\ref{sigma-LP-fact}).

We take $m_c = 1.5$~GeV. We use the CTEQ6M parton distribution functions
and the two-loop expression for $\alpha_s$, with $n_f = 5$
quark flavors and $\Lambda_{\rm QCD}^{(5)} = 226$~MeV. We set the
renormalization scale $\mu_r$ and the factorization scale
$\mu_f$ for the both parton distribution functions and the
FFs to be $m_T = \sqrt{p_T^2 +4 m_c^2}$. In order to resum leading
logarithms of $p_T^2/m_c^2$, we evolve the FFs from the scale $\mu_0=2
m_c$ to the scale $\mu_f=m_T\approx p_T$. We take
the NRQCD factorization scale to be $\mu_\Lambda=m_c$. In the
calculation of the PPCSs and the evolution of the FFs, we take $n_f=3$
active-quark flavors. That is, we ignore contributions from virtual or
initial heavy quarks.

\subsection{Parton production cross sections}

The PPCSs through order $\alpha_s^3$ were computed in the modified
minimal-subtraction ($\overline{\rm MS}$) scheme in
Refs.~\cite{Aversa:1988vb, Jager:2002xm}. We carry out numerical
computations of the PPCSs through order $\alpha_s^3$ by making use of
the computer code that was written by the authors of
Ref.~\cite{Aversa:1988vb}.

The PPCSs are computed as a function of $p_T$, $y$, and $z =
p^+/p_i^+=p_T/p_{iT}$, where $p_T$ is the transverse momentum of the $Q \bar
Q$ pair, $y$ is the rapidity of the $Q \bar Q$ pair in the hadron
center-of-momentum frame, and $p_{iT}$ is the transverse momentum of the
specific parton that is produced in the semi-inclusive partonic
scattering process. Here, we have written $z$ in terms of the transverse
momenta by using the fact that, in the LP approximation, one can ignore
the invariant mass of the $Q \bar Q$ pair. The maximum value of $p_{iT}$ is
kinematically constrained, and, so, the PPCSs vanish for $z \leq z_0 =
\frac{p_T}{\sqrt{s}} (e^{+y}+ e^{-y})$, where $\sqrt{s}$ is the
center-of-mass energy.

\subsection{Fragmentation functions}

In this paper we take into account FFs through order $\alpha_s^2$, which
are available for fragmentation of both gluons and quarks into polarized
and unpolarized $Q\bar Q$ pairs. A summary of FFs that we use in our
calculation can be found in Ref.~\cite{Ma:2013yla} and
Ref.~\cite{Ma:2015yka} for unpolarized and polarized $Q\bar Q$ pairs,
respectively. We give a detailed description below of the sources of
these FFs.

The gluon FF $D_{g \to Q \bar Q(n)}$ for $n={}^3S_1^{[8]}$ was
calculated for both unpolarized and polarized final states at
order $\alpha_s$ (LO) in Ref.~\cite{Braaten:1994kd} and at order
$\alpha_s^2$ (NLO) in Refs.~\cite{Braaten:2000pc, Ma:2013yla}. The gluon
FF for $n={}^1S_0^{[8]}$ was calculated at order $\alpha_s^2$ (LO) in
Refs.~\cite{Braaten:1996rp, Bodwin:2012xc}. The gluon FFs for
$n={}^3P_J^{[8]}$ were calculated at order $\alpha_s^2$ (LO) in
Refs.~\cite{Bodwin:2012xc, Braaten:1994kd} for unpolarized final states
and in Ref.~\cite{Ma:2015yka} for polarized final states. The gluon FFs
for $n={}^3P_J^{[1]}$ were calculated at order $\alpha_s^2$ (LO) in
Ref.~\cite{Braaten:1994kd} for unpolarized final states and in
Refs.~\cite{Cho:1994gb, Ma:2015yka} for polarized final states.

The situation for quark FFs $D_{q \to Q \bar Q(n)}$ with $n$ an $S$-wave
state is rather complicated, as there are several independent
calculations, some of which do not agree. Let us distinguish three
cases: (i) $q\neq Q$, in which case, $n={}^3S_1^{[8]}$; (ii) $q=Q$ and
$n={}^3S_1^{[8]}$; (iii) $q=Q$ and $n={}^3S_1^{[1]}$. The quark FF for
case~(i) for an unpolarized final state was calculated at order
$\alpha_s^2$ (LO) in Refs.~\cite{Ma:1995vi,Ma:2013yla,Bodwin:2014bia},
whose results all agree. The quark FF for case~(i) for a polarized final
state was calculated at order $\alpha_s^2$ (LO) in
Refs.~\cite{hong-zhang,Bodwin:2014bia,Ma:2015yka}. The results in
Refs.~\cite{Bodwin:2014bia,Ma:2015yka} agree with each other, but
disagree with the result in Ref.~\cite{hong-zhang}. The results in
Refs.~\cite{Bodwin:2014bia,Ma:2015yka} have since been confirmed by the
author of Ref.~\cite{hong-zhang}. The quark FF for case~(ii) for an
unpolarized final state was calculated at order $\alpha_s^2$ (LO) in
Refs.~\cite{Ma:2013yla,Yuan:1994hn,Bodwin:2014bia,Ma:1995vi}. The
results in Refs.~\cite{Ma:2013yla,Bodwin:2014bia} agree with each other
and disagree with the results in
Refs.~\cite{Yuan:1994hn,Ma:1995vi}. We use the results in
Refs.~\cite{Ma:2013yla,Bodwin:2014bia} in this paper. The quark FF for
case~(ii) for a polarized final state was calculated at order
$\alpha_s^2$ (LO) in Refs.~\cite{hong-zhang,Bodwin:2014bia,Ma:2015yka},
whose results agree. The quark FF for case~(iii) for an unpolarized
final state was calculated at order $\alpha_s^2$ (LO) in
Refs.~\cite{Braaten:1993mp,Ma:2013yla,Bodwin:2014bia}, whose results
agree. The quark FF for case~(iii) for a polarized final state was
calculated at order $\alpha_s^2$ (LO) in
Refs.~\cite{hong-zhang,Bodwin:2014bia,Ma:2015yka}, whose results agree.

The quark FFs $D_{Q \to Q \bar Q(n)}$ for $n={}^3P_J^{[1]}$ and
$n={}^3P_J^{[8]}$ were calculated for the unpolarized and polarized
cases at order $\alpha_s^2$ (LO) in Ref.~\cite{Ma:1995vi}.

The gluon FF $D_{g \to Q \bar Q({}^3S_1^{[1]})}$ was calculated at order
$\alpha_s^3$ (LO) in Refs.~\cite{Braaten:1993rw,Braaten:1995cj}.
Because the contributions to the FF in the ${}^3S_1^{[1]}$ channel begin
at order $\alpha_s^3$, we do not include them in our LP-fragmentation
calculations. However, we do use the LO FF for the ${}^3S_1^{[1]}$
channel to estimate the size of the uncalculated LP-fragmentation
contributions for that channel.

\subsection{DGLAP equation\label{sec:DGLAP}}

At leading order in $\alpha_s$, the DGLAP equation is given by
\cite{Gribov:1972ri, Lipatov:1974qm, Dokshitzer:1977sg,
Altarelli:1977zs}
\begin{equation}
\frac{d}{d \log \mu_f^2}
\begin{pmatrix}
D_S (\mu_f) \\ D_g(\mu_f)
\end{pmatrix}
=
\frac{\alpha_s(\mu_f)}{2 \pi}
\begin{pmatrix}
P_{qq} & ~2 n_f P_{gq} \\
P_{qg} & ~P_{gg}
\end{pmatrix}
\otimes
\begin{pmatrix}
D_S (\mu_f)\\ D_g(\mu_f)
\end{pmatrix},
\label{eq:DGLAP}
\end{equation}
where $D_g = D_{g \to Q \bar Q(n)}$, $D_S = \sum_f [ D_{q_f \to Q \bar
Q(n)} + D_{\bar q_f \to Q \bar Q(n)}]$, $f$ is the light-quark or
light-antiquark flavor, the $P_{ij}$ are the splitting functions for
the FFs, and $n_f$ is the number of active light-quark flavors.
The symbol $\otimes$ represents the convolution
\begin{equation}
(f \otimes g) (z) = \int_0^1 dx \int_0^1 dy
f(x) g(y) \delta(xy-z)
= \int_z^1 \frac{dx}{x} f(z/x) g(x)
= \int_z^1 \frac{dx}{x} f(x) g(z/x).
\end{equation}
The splitting functions are given by
\begin{subequations}
\begin{eqnarray}
P_{gg} (z) &=& 2 C_A
\left[ \frac{z}{(1-z)_+} + \frac{1-z}{z} + z (1-z) +
\frac{b_0}{12} \delta (1-z) \right],
\\
P_{gq} (z) &=& C_F \frac{1+(1-z)^2}{z},
\\
P_{qg} (z) &=& T_F [z^2 + (1-z)^2],
\\
P_{qq} (z) &=& C_F \left[ \frac{1+z^2}{(1-z)_+} + \frac{3}{2} \delta(1-z)
\right],
\end{eqnarray}
\end{subequations}
where
\begin{subequations}
\begin{eqnarray}
C_A &=&N_c
,
\\
C_F&=&\frac{N_c^2-1}{2N_c}
,
\\
T_F&=&\frac{1}{2}
,
\\
b_0 &=& \frac{11}{3} N_c - \frac{2}{3} n_f,
\end{eqnarray}
\end{subequations}
and $N_c=3$ is the number of colors.

As is well known, an analytic solution to Eq.~(\ref{eq:DGLAP}) can be
obtained in Mellin space. The Mellin transform of a function $f$ is
defined by
\begin{equation}
\tilde f (N) = ({\cal M} f) (N) = \int_0^1 dz\, z^{N-1} f(z),
\end{equation}
where we use a tilde ($\tilde {\phantom{f}}$) to denote objects in
Mellin space. The Mellin transform of the convolution in
Eq.~(\ref{eq:DGLAP}) is an ordinary product:
\begin{equation}
[{\cal M} (f \otimes g)] (N) =
({\cal M} f) (N) \times ({\cal M} g) (N) .
\label{eq:Mellin_conv}
\end{equation}
Hence, Eq.~(\ref{eq:DGLAP}) can be diagonalized by taking the Mellin
transform. Using the one-loop evolution of $\alpha_s$
\begin{equation}
\frac{d}{d \log \mu_f^2}
=
- \frac{b_0}{4\pi}
\,\alpha_s^2 (\mu_f)\,
\frac{d}{d \alpha_s (\mu_f)} ,
\end{equation}
one obtains the following solution of the DGLAP equation:
\begin{equation}
\begin{pmatrix}
\tilde D_S (N,\mu_f) 
\\ \tilde D_g (N,\mu_f)
\end{pmatrix}
=
\left[
M_+
\left(\frac{\alpha_s(\mu_0)}{\alpha_s(\mu_f)} \right)^{2 \tilde P^+/b_0}
+ M_-
\left(\frac{\alpha_s(\mu_0)}{\alpha_s(\mu_f)} \right)^{2 \tilde P^-/b_0}
\right]
\begin{pmatrix}
\tilde D_S (N,\mu_0) \\ \tilde D_g (N,\mu_0)
\label{mellin-space-solution}
\end{pmatrix} ,
\end{equation}
where
\begin{equation}
M_{\pm} =
\pm \frac{1}{\tilde P^+ - \tilde P^-}
\begin{pmatrix}
\tilde P_{qq}-\tilde P^\mp & 2 n_f \tilde P_{gq} \\
\tilde P_{qg} & \tilde P_{gg} - \tilde P^\mp
\end{pmatrix},
\end{equation}
and
\begin{equation}
\tilde P^\pm = \frac{1}{2} \left[\tilde P_{gg} + \tilde P_{qq} \pm
{\textstyle
\sqrt{(\tilde P_{gg} - \tilde P_{qq})^2 + 8 n_f \tilde P_{qg} \tilde P_{gq}}}
\right].
\end{equation}
The evolved FFs in $z$-space can be obtained from
Eq.~(\ref{mellin-space-solution}) by applying the inverse Mellin
transform
\begin{equation}
D_{i \to Q \bar Q(n)} (z,\mu_f) =
\frac{1}{2 \pi i} \int_{c-i \infty}^{c+i \infty} dN \, z^{-N}
\tilde D_{i \to Q \bar Q(n)} (N,\mu_f),
\label{inverse-mellin-tx}
\end{equation}
where the real number $c$ is chosen so that the integral over $N$
follows a contour that lies to the right of all of the poles of $\tilde{D}_{i
\to Q \bar Q(n)} (N,\mu_f)$.

We resum the leading logarithms of $p_T^2/m_c^2$ by choosing the
evolution scales $\mu_0=2m_c$ and $\mu_f=m_T\approx p_T$. In this paper,
we compute the integral over $N$ numerically, using analytic
expressions for the Mellin transforms of the FFs at
the scale $\mu_0=2m_c$.

There is a difficulty in numerical computation of the inverse Mellin
transform in Eq.~(\ref{inverse-mellin-tx}) near $z = 1$. For $z \ll 1$,
the factor $z^{-N}$ causes the integrand to vanish quickly at large
$|N|$ and the integral over $N$ converges. On the other hand, when $z =
1$, the convergence of the integral depends solely on the behavior of
the Mellin-space FF $\tilde D_{i \to Q \bar Q(n)} (N,\mu_f)$ at
large $|N|$. Since $\tilde{P}_{gg}$ and $\tilde{P}_{qq}$ behave
asymptotically as negative constants times $\log |N|$, while
$\tilde{P}_{gq}$ and $\tilde{P}_{qg}$ vanish asymptotically as inverse
powers of $N$, the coefficients of $M_+$ and $M_-$ in
Eq.~(\ref{inverse-mellin-tx}) damp the integral when
$\alpha_s(\mu_f)\ll \alpha_s(\mu_0)$. However, the integrals do
not converge at $z=1$ unless $\mu_f$ is quite large in comparison
with $\mu_0$. In fact, as $\mu_f$ approaches $\mu_0$, the evolved
FFs approximate the initial FFs, which, in some cases, are
distributions at $z=1$. We deal with this problem by rearranging the
convolutions of the FFs and the PPCSs so as to treat the singular
behavior of the FFs at $z=1$ analytically. The details of the method are
given in the Appendix.

\section{Results for Combined LP and NLO Short-Distance
Coefficients\label{sec:LP+NLO}}

\begin{figure}
\epsfig{file=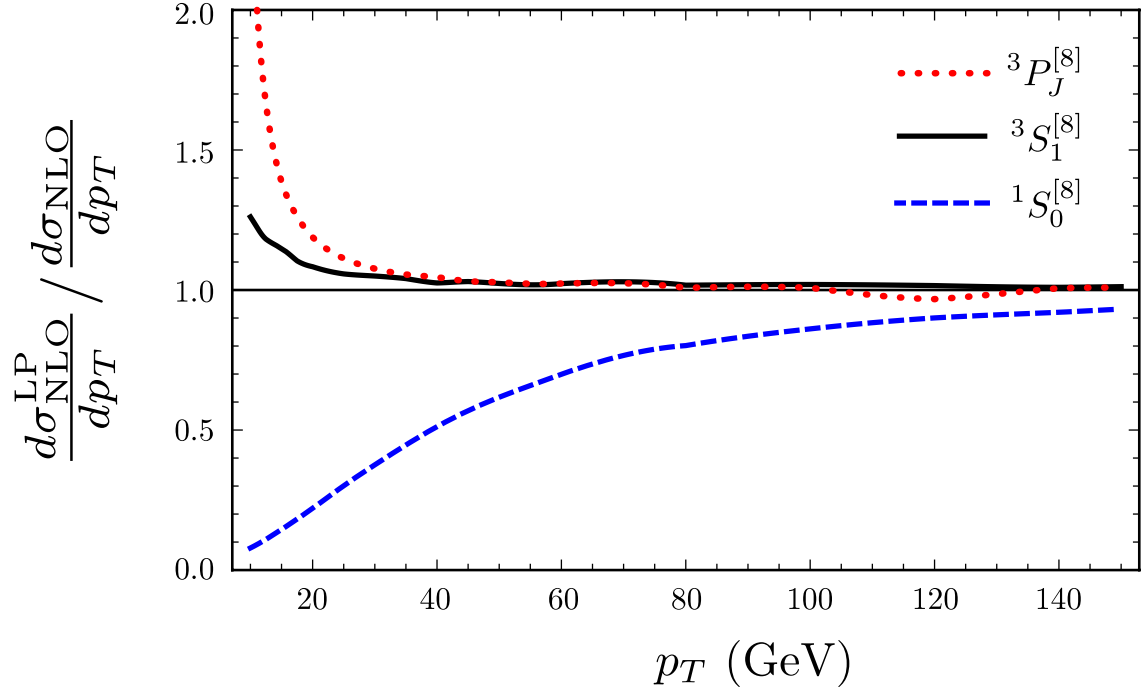,width=15cm}
\caption{\label{fig:ratio_NLO1}%
The ratio
$(d\sigma^{\textrm{LP}}_{\rm NLO}/dp_T) / (d\sigma_{\rm NLO}/dp_T)$
for the ${}^1S_0^{[8]}$, ${}^3P_J^{[8]}$, and ${}^3S_1^{[8]}$ channels
in the process $p p \to H + X$ at $\sqrt{s}=7$~TeV and $|y|<1.2$.
}
\end{figure}

\begin{figure}
\epsfig{file=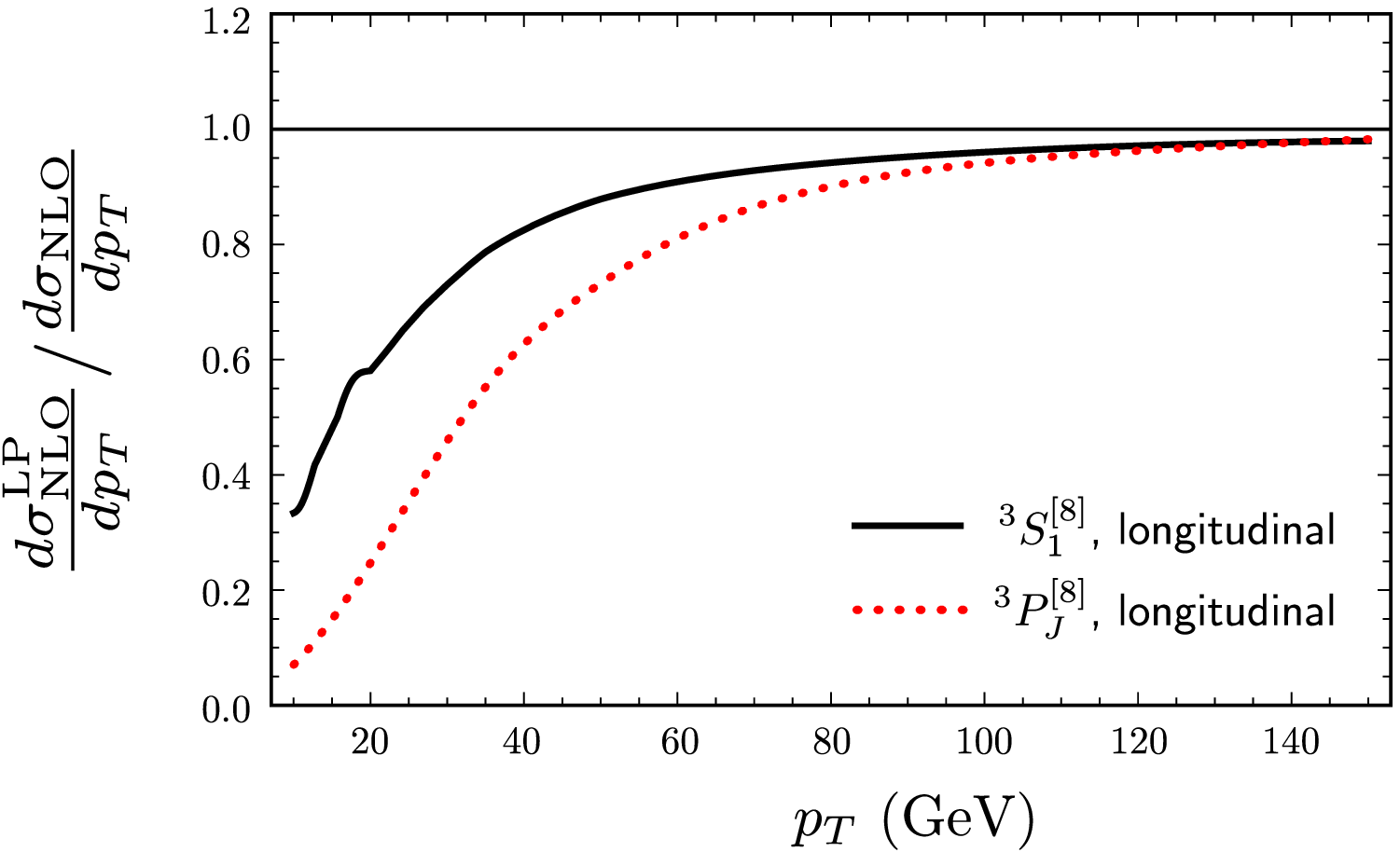,width=15cm}
\caption{\label{fig:ratio_NLO2}%
The ratio
$(d\sigma^{\textrm{LP}}_{\rm NLO}/dp_T) / (d\sigma_{\rm NLO}/dp_T)$
for the polarized ${}^3P_J^{[8]}$ and ${}^3S_1^{[8]}$ channels with longitudinal
final states
in the process $p p \to H + X$ at $\sqrt{s}=7$~TeV and $|y|<1.2$.
}
\end{figure}

\begin{figure}
\epsfig{file=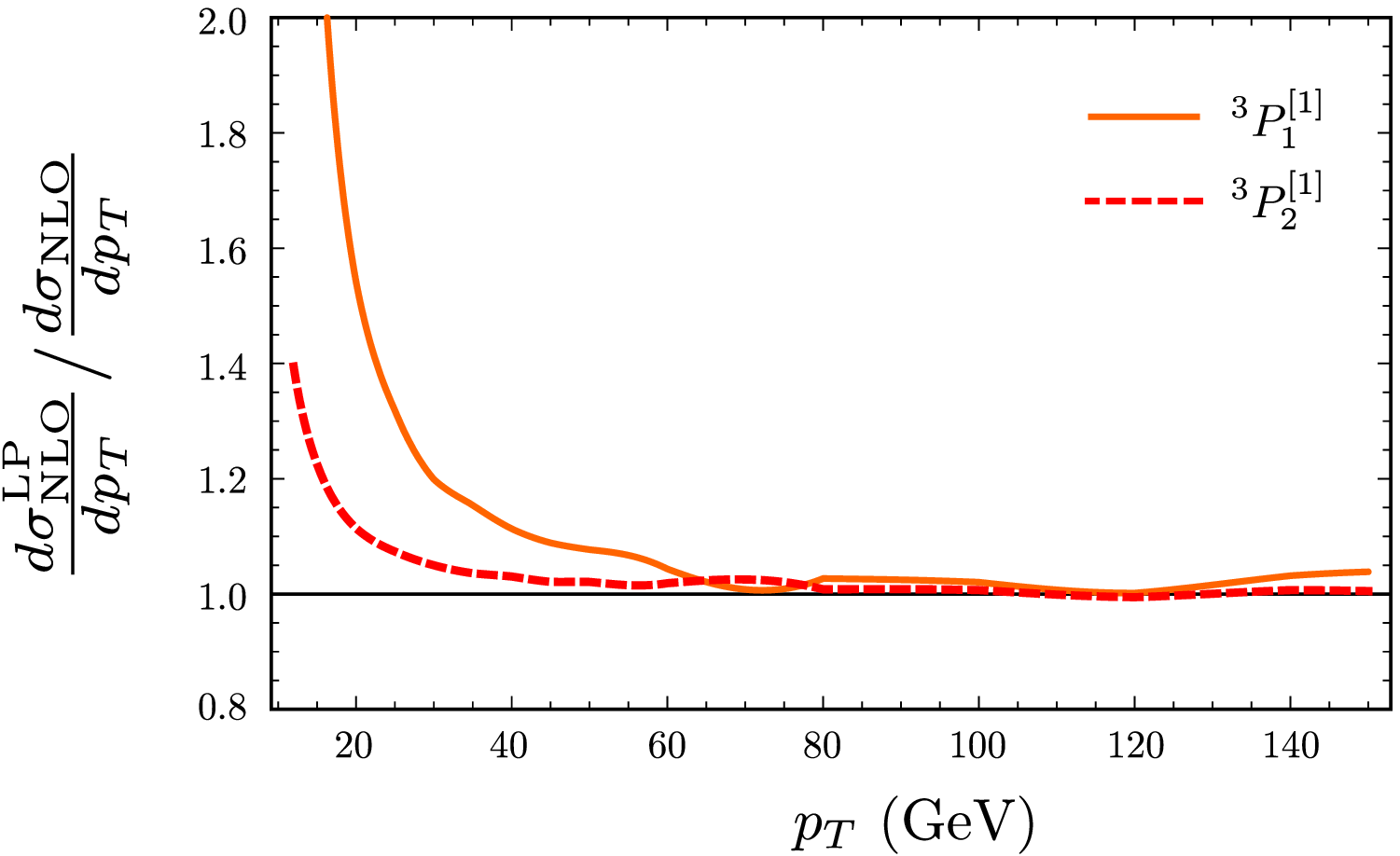,width=15cm}
\caption{\label{fig:ratio_NLO3}%
The ratio
$(d\sigma^{\textrm{LP}}_{\rm NLO}/dp_T) / (d\sigma_{\rm NLO}/dp_T)$
for the ${}^3P_1^{[1]}$ and ${}^3P_2^{[1]}$ channels
in the process $p p \to H + X$ at $\sqrt{s}=7$~TeV and $|y|<1.2$.
}
\end{figure}

\begin{figure}
\epsfig{file=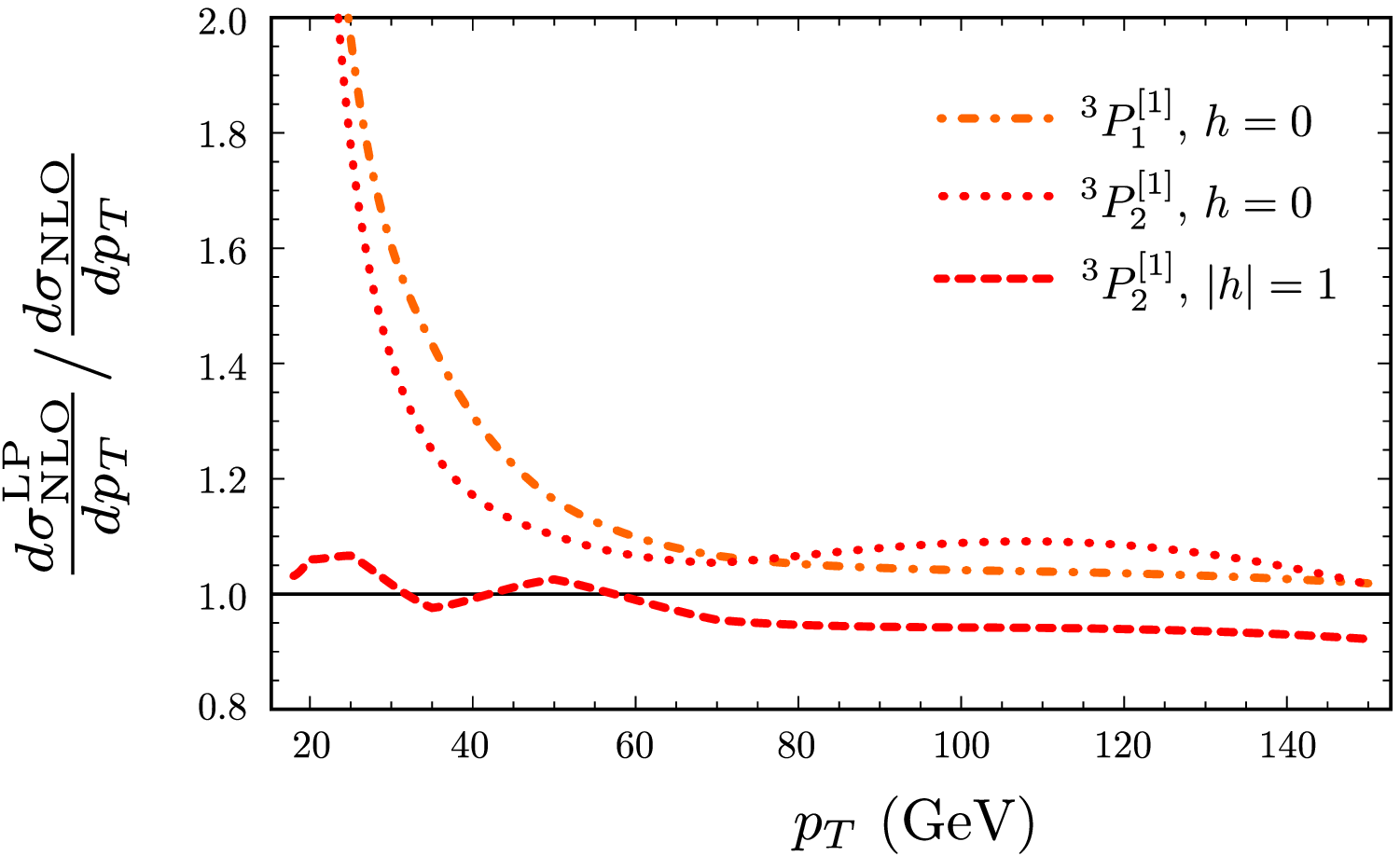,width=15cm}
\caption{\label{fig:ratio_NLO4}%
The ratio
$(d\sigma^{\textrm{LP}}_{\rm NLO}/dp_T) / (d\sigma_{\rm NLO}/dp_T)$
for the polarized ${}^3P_1^{[1]}$ and ${}^3P_2^{[1]}$ channels
in the process $p p \to H + X$ at $\sqrt{s}=7$~TeV and $|y|<1.2$.
$h$ is the helicity of the $Q \bar Q$ pair in the final state.
}
\end{figure}

Now we use Eq.~(\ref{LP+NLO}) to combine results for the LP SDCs,
computed as described in Sec.~\ref{sec:LP-SDCs}, with the SDCs through
NLO in $\alpha_s$. For the latter, we make use of the computations in
Refs.~\cite{Ma:2010jj,Ma:2010yw,Chao:2012iv}, taking the values of the
parton distributions, $m_c$, $\alpha_s$, $\mu_r$, $\mu_f$,
$\mu_\Lambda$, and $n_f$ that are specified at the start of
Sec.~\ref{sec:LP-SDCs}.\footnote{In order to improve computational
efficiency, we have omitted in the calculation of $d\sigma_{\rm
NLO}/dp_T$ contributions from processes that are initiated by two
light quarks, two light-antiquarks, or a light quark and a light
antiquark, where the two initial partons can have different flavors. We
use the generic expression $qq$ to denote these light-quark/antiquark
initial states. The $qq$-initiated contributions are small in comparison
to the sum of the $qg$- and $gg$-initiated contributions because the $q$
and $\bar q$ partonic fluxes are small in comparison to the $g$ partonic
flux. As $p_T$ increases, the sizes of the $q$ and $\bar q$ partonic
fluxes increase relative to the size of the $g$ partonic flux because
larger values of the parton momentum fractions are emphasized. At large
values of $p_T$, $d\sigma_{\rm NLO}/dp_T$ is well approximated by
$d\sigma_{\rm NLO}^{\rm LP}/dp_T$. Therefore, we adopt the following
computational strategy. In order to match what was done in the NLO
calculation, we omit the $qq$-initiated contributions in computing
$d\sigma_{\rm NLO}^{\rm LP}/dp_T$ in Eq.~(\ref{LP+NLO}). However, we
take the $qq$-initiated contributions into account at large $p_T$, where
they can be more important, by including them in the computation of
$d\sigma^{\rm LP}/dp_T$ in Eq.~(\ref{LP+NLO}). Since each
$qq$-initiated process that produces a given $Q\bar Q$ channel contains
an LP fragmentation contribution at the leading nontrivial order in
$\alpha_s$, we can use LP fragmentation results to estimate the sizes of
the $qq$-initiated contributions. These estimates indicate that
$qq$-initiated contributions produce the largest fractional correction
in the longitudinally polarized ${}^3S_1^{[8]}$ channel, in which they
grow to about $5\%$ of the total at $p_T=100$~GeV. Hence, we expect any
errors that result from the omission of the $qq$-initiated processes in
the NLO calculations to be much less than $5\%$.}

We first compare our results for $d \sigma^{\rm LP}_{\rm NLO}/d p_T$
with $d \sigma_{\rm NLO}/d p_T$, the fixed-order SDC accurate through
NLO. Figures~\ref{fig:ratio_NLO1}--\ref{fig:ratio_NLO4}
show the ratios $(d \sigma^{\rm LP}_{\rm NLO}/d p_T)/
(d \sigma_{\rm NLO}/d p_T)$ for the polarized and unpolarized final
states in the process $p p \to H + X$ at $\sqrt{s}=7$~TeV and $|y|<1.2$.

In Fig.~\ref{fig:ratio_NLO1}, we show the ratios $(d \sigma^{\rm LP}_{\rm
NLO}/d p_T)/ (d \sigma_{\rm NLO}/d p_T)$ for unpolarized final
states in the ${}^3S_1^{[8]}, {}^1S_0^{[8]}$, and ${}^3P_J^{[8]}$
channels. As $p_T$ increases, the ratios for the ${}^3S_1^{[8]}$ and
${}^3P_J^{[8]}$ channels quickly approach unity because the
LP-fragmentation contribution dominates the SDCs. This approach to unity
is slower for the ${}^1S_0^{[8]}$ channel because the FF for the
${}^1S_0^{[8]}$ channel does not receive enhancements near $z=1$ from
a Dirac $\delta$ function or plus distributions that are the remnants
of soft divergences that cancel between real and virtual gluon-emission
processes.\footnote{ It was shown in Ref.~\cite{Ma:2014svb} that the
ratio $(d \sigma^{\rm LP}_{\rm NLO}/d p_T+d \sigma^{\rm NLP}_{\rm NLO}/d
p_T)/ (d \sigma_{\rm NLO}/d p_T)$, which takes into account both the LP
and NLP contributions, approaches unity much faster for the
${}^1S_0^{[8]}$ channel than does the ratio $(d \sigma^{\rm LP}_{\rm
NLO}/d p_T)/ (d \sigma_{\rm NLO}/d p_T)$.} At small $p_T$, the ratio
$(d \sigma^{\rm LP}_{\rm NLO}/d p_T)/ (d \sigma_{\rm NLO}/d p_T)$ is
larger for the ${}^3P_J^{[8]}$ channel than for the ${}^3S_1^{[8]}$ and
${}^1S_0^{[8]}$ channels because, for the ${}^3P_J^{[8]}$ channel, the
LO and NLO contributions in the denominator tend to cancel.

In Fig.~\ref{fig:ratio_NLO2}, we show the ratios $(d \sigma^{\rm LP}_{\rm
NLO}/d p_T)/ (d \sigma_{\rm NLO}/d p_T)$ for longitudinal final
states in the ${}^3S_1^{[8]}$ and ${}^3P_J^{[8]}$ channels. The approach
of each of these ratios to unity is slow. As was the case for the ratio
of cross sections in the ${}^1S_0^{[8]}$ channel, the slow approach to
unity is a consequence of the fact that the FFs are not enhanced near
$z=1$ by a Dirac $\delta$ function or plus distributions.

In Fig.~\ref{fig:ratio_NLO3}, we show the ratios $(d \sigma^{\rm LP}_{\rm
NLO}/d p_T)/ (d \sigma_{\rm NLO}/d p_T)$  for unpolarized final states
in the color-singlet $P$-wave channels. We show the ratios for the
polarized final states in Fig.~\ref{fig:ratio_NLO4}. Here, $h$ is the
helicity of the $Q \bar Q$ pair in the final state. The behaviors are
similar to those for the ${}^3P_J^{[8]}$ channel, except for the case
of $J=2$ with $|h|=1$, for which the ratio 
$(d \sigma^{\rm LP}_{\rm NLO}/d p_T)/ (d
\sigma_{\rm NLO}/d p_T)$ is almost constant. We note that the
deviation of $(d \sigma^{\rm LP}_{\rm NLO}/d p_T)/ (d \sigma_{\rm NLO}/d
p_T)$ from unity at large $p_T$ is of the same relative size as the
statistical uncertainty in the NLO calculation.

\begin{figure}
\epsfig{file=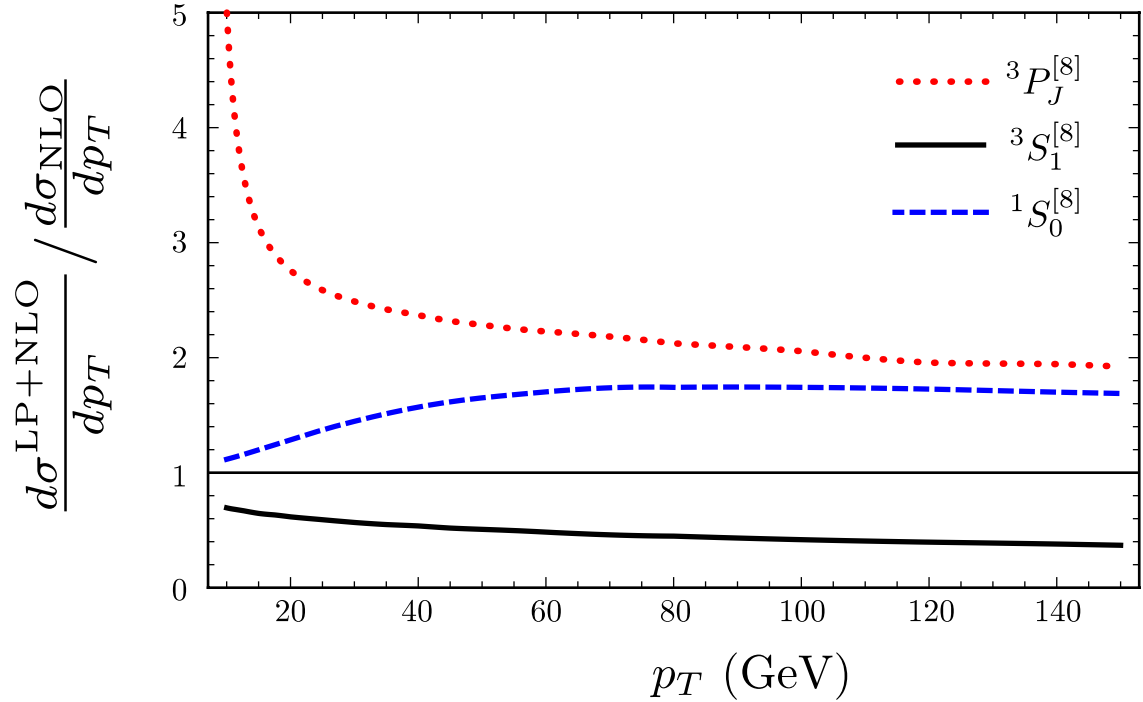,width=15cm}
\caption{\label{fig:ratio_NLO5}%
The ratio $(d\sigma^{{\rm LP+NLO}}/dp_T) / (d\sigma_{\rm NLO}/dp_T)$ for
the ${}^1S_0^{[8]}$, ${}^3P_J^{[8]}$, and ${}^3S_1^{[8]}$ channels with
unpolarized final states in the process $p p \to H + X$ at
$\sqrt{s}=7$~TeV and $|y|<1.2$.
}
\end{figure}

\begin{figure}
\epsfig{file=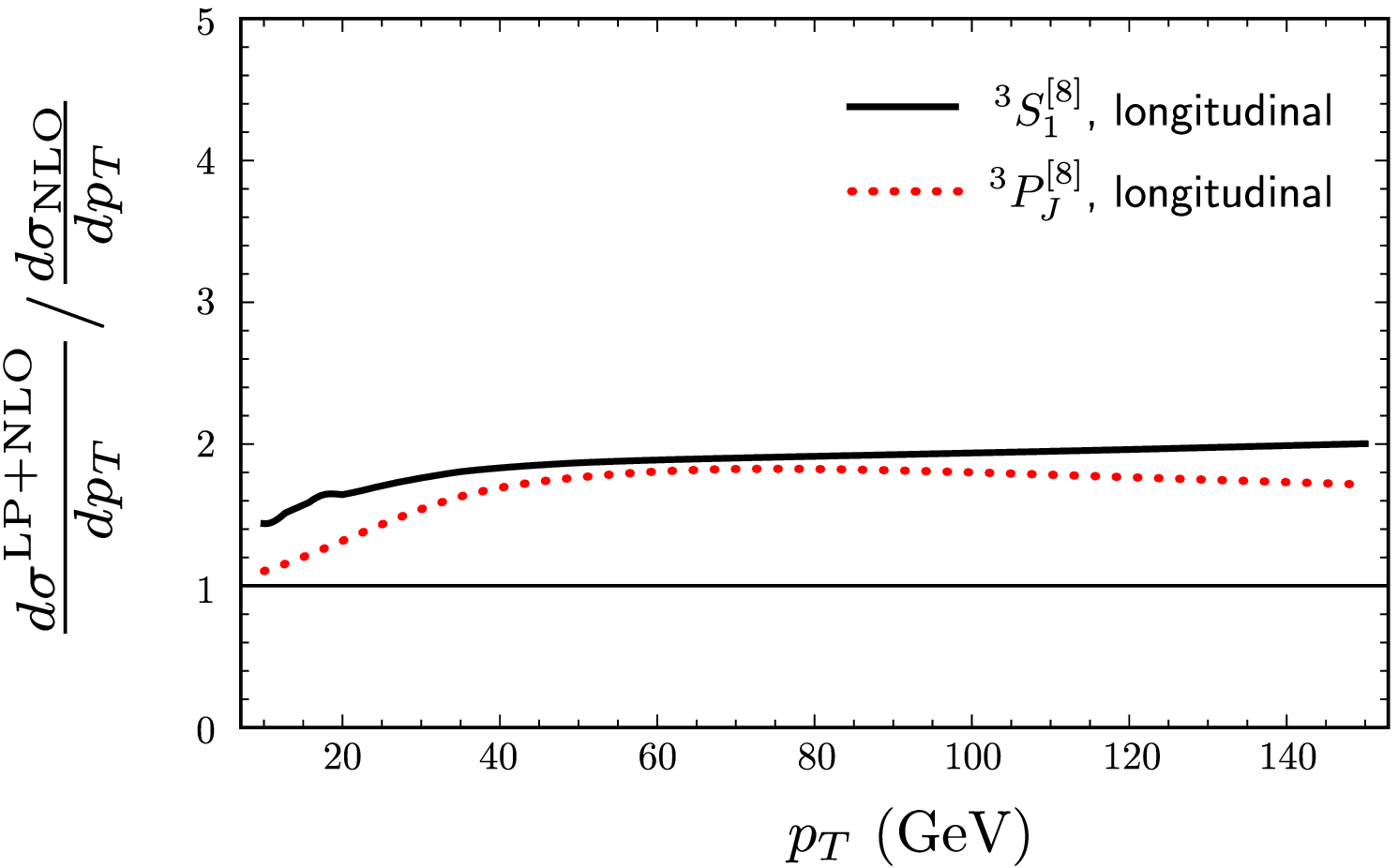,width=15cm}
\caption{\label{fig:ratio_NLO6}%
The ratio $(d\sigma^{{\rm LP+NLO}}/dp_T) / (d\sigma_{\rm NLO}/dp_T)$ for
the polarized ${}^3P_J^{[8]}$ and ${}^3S_1^{[8]}$ channels with
longitudinal final states in the process $p p \to H + X$ at
$\sqrt{s}=7$~TeV and $|y|<1.2$.
}
\end{figure}

\begin{figure}
\epsfig{file=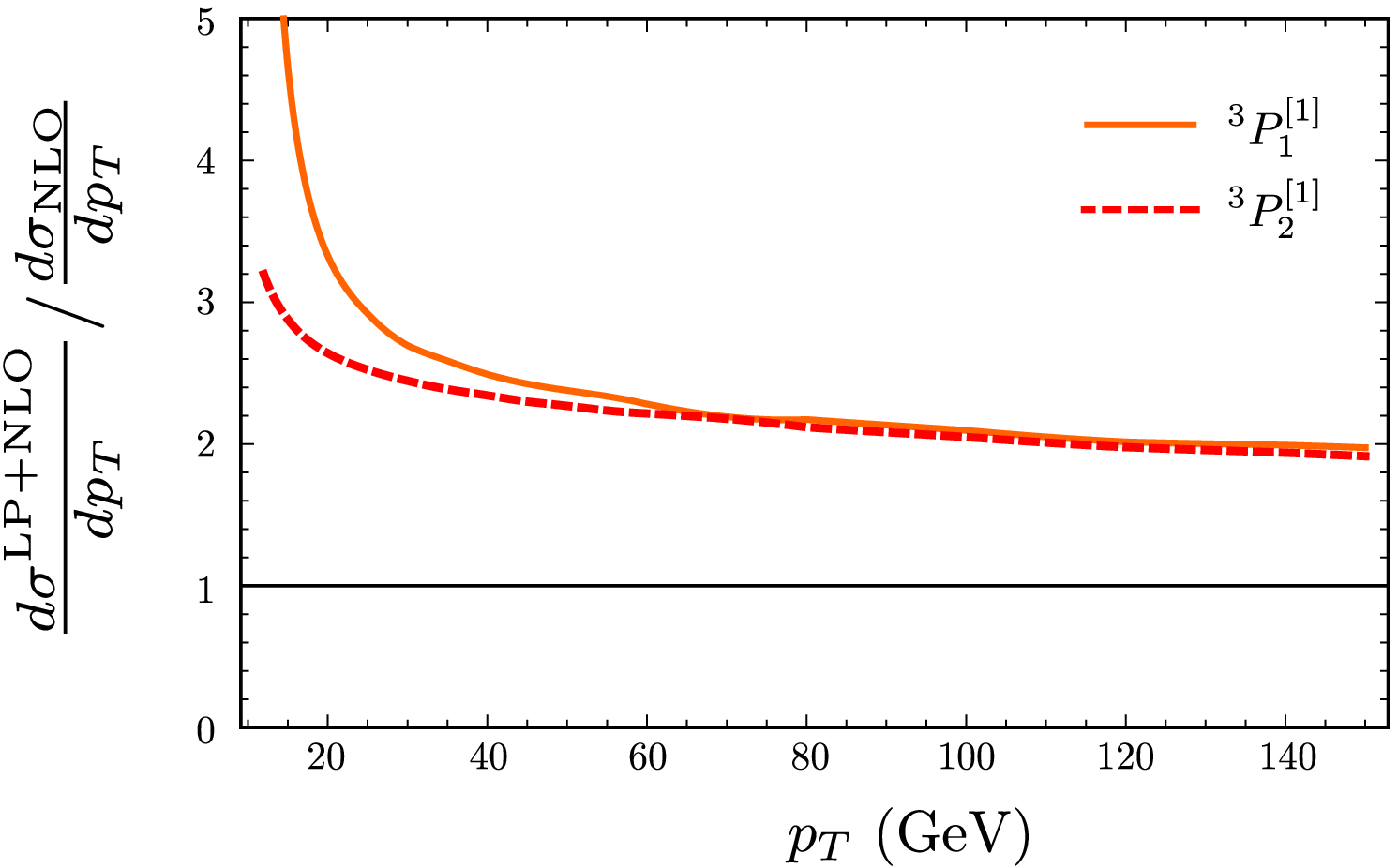,width=15cm}
\caption{\label{fig:ratio_NLO7}%
The ratio $(d\sigma^{{\rm LP+NLO}}/dp_T) / (d\sigma_{\rm NLO}/dp_T)$ for
the ${}^3P_1^{[1]}$ and ${}^3P_2^{[1]}$ channels with unpolarized final
states in the process $p p \to H + X$ at $\sqrt{s}=7$~TeV and $|y|<1.2$.
}
\end{figure}

\begin{figure}
\epsfig{file=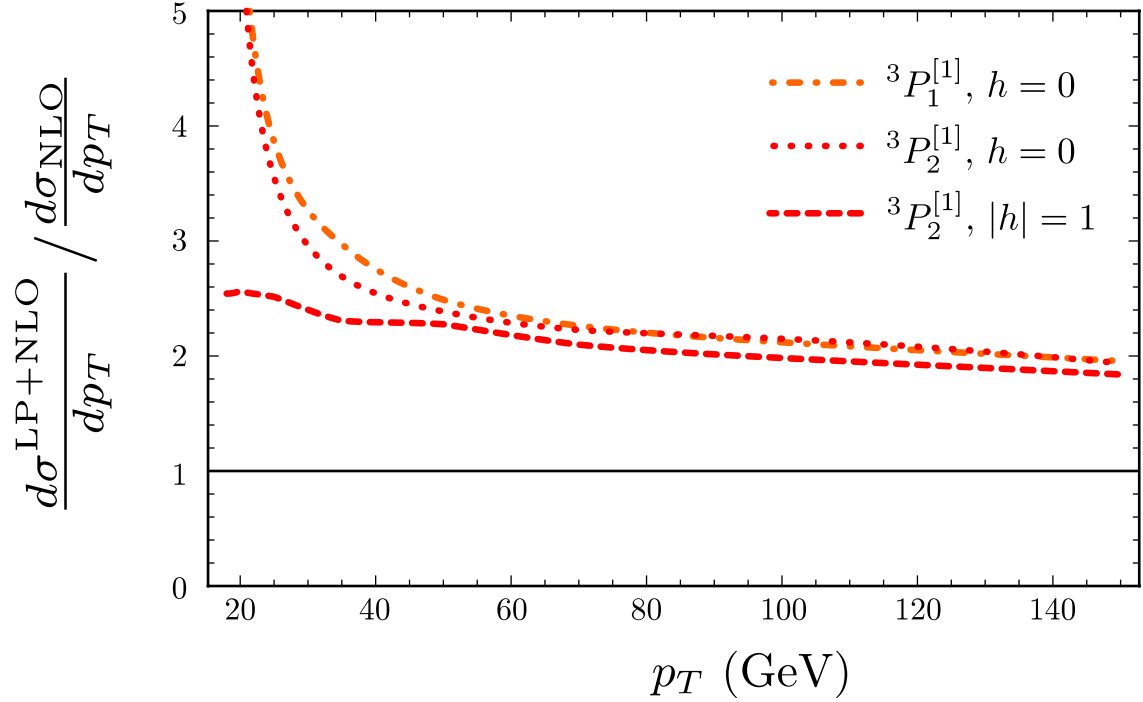,width=15cm}
\caption{\label{fig:ratio_NLO8}%
The ratio $(d\sigma^{{\rm LP+NLO}}/dp_T) / (d\sigma_{\rm NLO}/dp_T)$ for
the ${}^3P_1^{[1]}$ and ${}^3P_2^{[1]}$ channels with polarized final
states in the process $p p \to H + X$ at $\sqrt{s}=7$~TeV and $|y|<1.2$.
}
\end{figure}

Next we compare $d \sigma^{\rm LP+NLO}_{\rm NLO}/d p_T$, the SDC
that includes both the fixed-order corrections through NLO and the
additional LP corrections, with $d \sigma_{\rm NLO}/d p_T$, the SDC that
includes fixed-order corrections through NLO. Specifically, we show the
ratios $(d \sigma^{\rm LP+NLO}_{\rm NLO}/d p_T)/ (d \sigma_{\rm NLO}/d
p_T)$ for the polarized and unpolarized final states in
Figs.~\ref{fig:ratio_NLO5}--\ref{fig:ratio_NLO8}. With the exception of
the ${}^3S_1^{[8]}$ channel, the additional LP-fragmentation
contributions are of the order of $100\%$ at large $p_T$. As we have
mentioned, because there is a partial cancellation between the LO and
the NLO contributions in the ${}^3P_J^{[8]}$ channel, the additional 
LP fragmentation corrections have a significant impact on the shape in that
channel. For the ${}^3S_1^{[8]}$ channel the additional
LP-fragmentation contributions are negative and only mildly alter the
shape.

As was pointed out in Ref.~\cite{Bodwin:2014gia}, the effects from the
all-orders resummation of logarithms of $p_T^2/m_c^2$ are small. In the
case of the ${}^3S_1^{[8]}$ channel, almost all of the effects of the
large logarithms are already accounted for in the NLO contribution. In
the cases of the ${}^1S_0^{[8]}$ and ${}^3P_J^{[8]}$ channels, the
all-orders resummations of logarithms shift the FFs by only about $2\%$
and $5\%$, respectively, at $p_T=52.7$~GeV because contributions from
the running of $\alpha_s$ and from the DGLAP splitting cancel. Hence,
almost all of the large additional LP corrections that we find arise
from nonlogarithmic contributions of order $\alpha_s^5$.

Finally, we discuss the LP-fragmentation contribution to the
${}^3S_1^{[1]}$ channel. Since the FF for this channel begins at order
$\alpha_s^3$, the ${}^3S_1^{[1]}$ channel receives an LP contribution
that begins at order $\alpha_s^5$ (NNLO). We do not include this
LP-fragmentation contribution in our analysis. However, we have
estimated its size by making use of the FF at order $\alpha_s^3$. At
$p_T = 10$~GeV, the LP contribution is about an order of magnitude
smaller than the fixed-order contribution through NLO. The LP
contribution reaches the same size as the fixed-order contribution
through NLO at around $p_T = 50$~GeV. Finally, when $p_T = 130$~GeV, the
LP contribution is almost an order of magnitude larger than the
fixed-order contribution through NLO. Although the LP-fragmentation
contribution can have a significant effect on the color-singlet
contribution at large $p_T$, its effect on the cross section is only of
the order of $1\%$ of the measured cross section at $p_T=130$~GeV.

\section{Fits of Cross-Section Predictions to Data
\label{sec:LDMEs}}

In this section we extract the color-octet LDMEs by fitting the
cross-section predictions that are based on the LP$+$NLO SDCs to the
measured cross sections. We use the resulting LDMEs to make predictions
for the prompt-$J/\psi$ polarization. In order to suppress possible
nonfactorizing contributions, we fit only to data for which $p_T$ is
greater than $ 3 m_H$, where $m_H$ is the quarkonium mass. Since the
shape of the $p_T$ distribution determines the LDMEs, it is crucial to
use the data at the highest $p_T$ values in the fits.

In the case of the direct $J/\psi$ cross section, we estimate the
theoretical uncertainties in the SDCs to be $25\%$ of the central
values. We arrived at these uncertainties by varying the factorization
scale $\mu_f$ and the renormalization scale $\mu_r$
independently between $\frac{1}{2} m_T$ and $2 m_T$. This $25\%$
uncertainty is also roughly the size of the uncertainty that one would
expect from uncalculated corrections of higher order in
$v$. In the cases of the cross sections of the excited charmonium
states, we take the uncertainties to be $30\%$ of the central values
because the $v^2$ for those states is larger than for the $J/\psi$.

\subsection{Production of $\bm{\psi(2S)}$}

We determine the three color-octet $\psi(2S)$ LDMEs by performing a 
least-$\chi^2$ fit to the CDF~\cite{Aaltonen:2009dm} and
CMS~\cite{Chatrchyan:2011kc, Khachatryan:2015rra} cross-section data. In
order to suppress possible nonfactorizing contributions, we use only
the data for which $p_T$ is greater than $11$~GeV. We ignore feeddown
contributions from decays of heavier quarkonia.

In the case of the color-singlet LDME, we take a value that was
determined in a  potential-model calculation~\cite{Eichten:1995ch}:
$\langle {\cal O}^{\psi(2S)} ({}^3S_1^{[1]}) \rangle = 0.76$~GeV${}^3$.
Different choices for the value of the color-singlet LDME would have
little effect on our results, as the contribution from the color-singlet
channel is much smaller than the theoretical uncertainties. In the
lowest $p_T$ bin that we consider for the CMS data that have $|y|<1.2$
($11\textrm{~GeV}< p_T < 12\textrm{~GeV}$), the contribution from the
color-singlet channel is only about $5\%$ of the cross section, and the
color-singlet contribution drops to $0.2\%$ in the highest $p_T$ bin
($75\textrm{~GeV}< p_T < 100\textrm{~GeV}$).

\begin{figure}
\epsfig{file=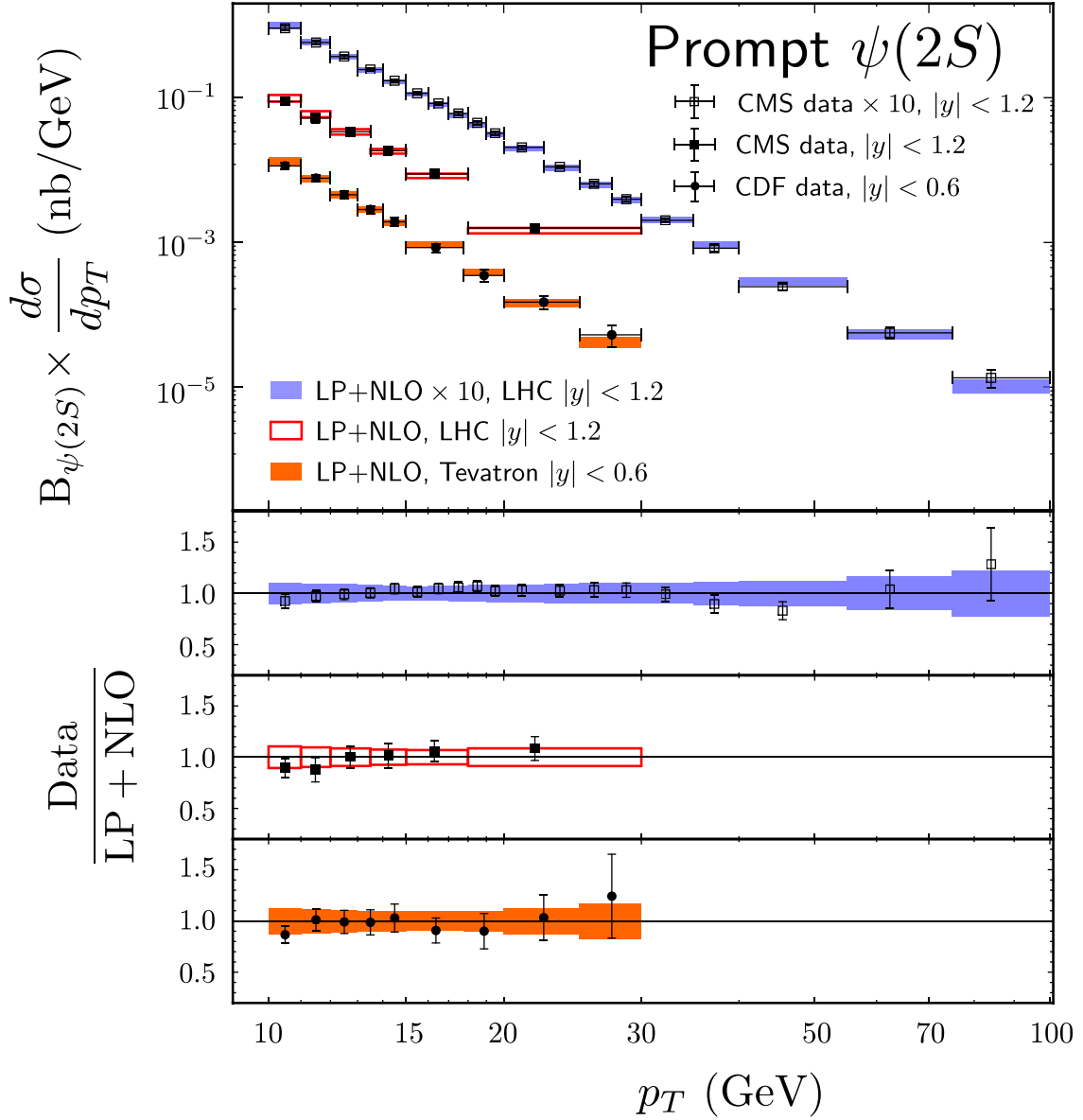,width=15cm}
\caption{\label{fig:psi2s_fit}%
The differential cross sections for prompt $\psi(2S)$
production at the Tevatron ($\sqrt{s}=1.96$~TeV) and
the LHC ($\sqrt{s}=7$~TeV).
$B_{\psi(2S)}={\rm Br}[\psi(2S)
\to \mu^+ \mu^-]$, where ${\rm Br}$ denotes a branching ratio.
}
\end{figure}

\begin{figure}
\epsfig{file=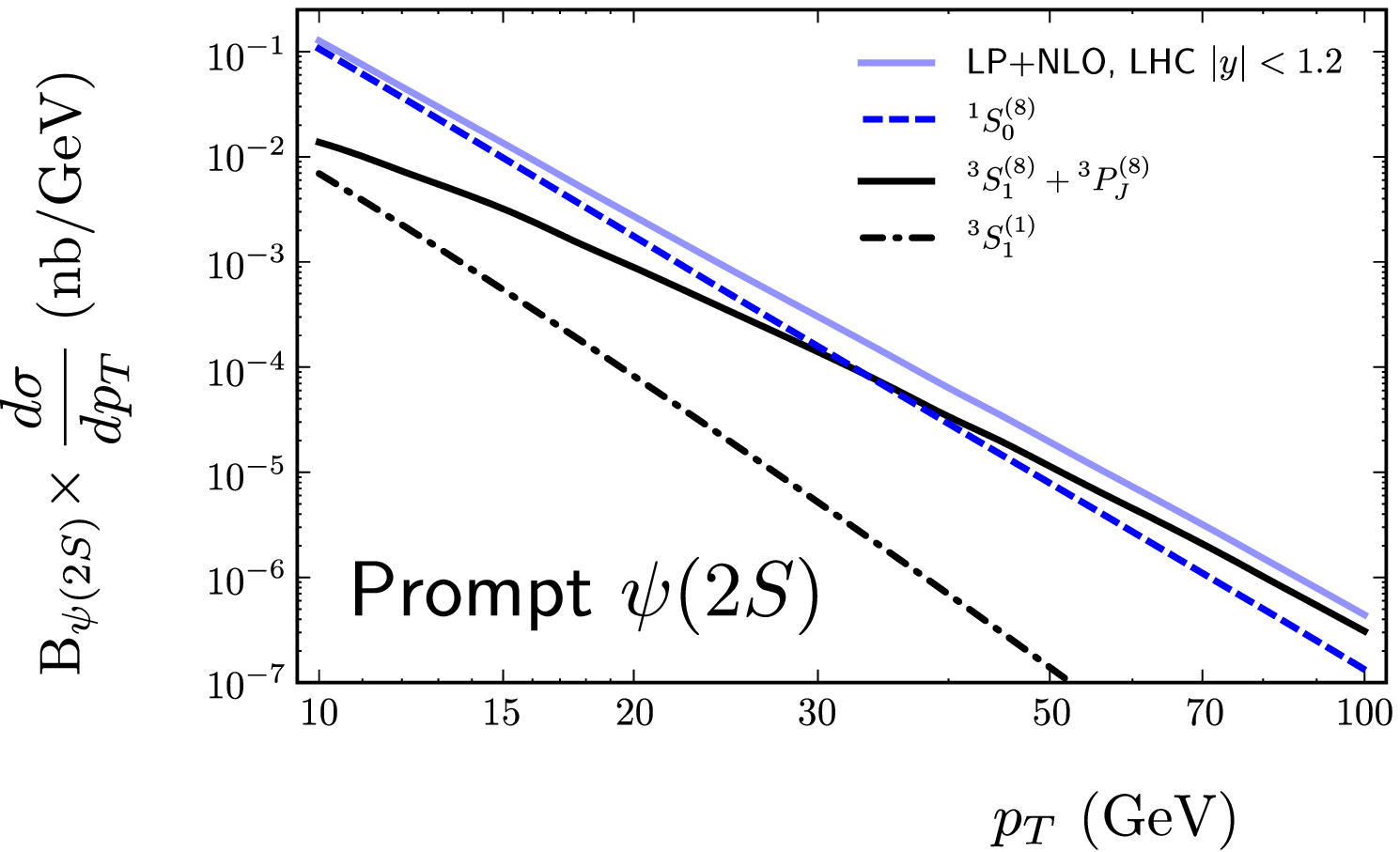,width=15cm}
\caption{\label{fig:psi2s_comp}%
Contributions of the individual channels to
the prompt $\psi(2S)$ differential cross section at the LHC
($\sqrt{s}=7$~TeV).
$B_{\psi(2S)}={\rm Br}[\psi(2S) \to \mu^+ \mu^-]$.
}
\end{figure}

The fitted LP$+$NLO cross section is compared with the data in
Fig.~\ref{fig:psi2s_fit}. The quality of the fit is quite good, with
$\chi^2/{\rm d.o.f.} = 1.71/29$. As can be seen from
Fig.~\ref{fig:psi2s_comp}, the cross section is dominated by the
${}^1S_0^{[8]}$ channel at moderate values of $p_T$, but not at large
values of $p_T$.  The concept of ${}^1S_0^{[8]}$ dominance has been
suggested previously in
Refs.~\cite{Chao:2012iv,Bodwin:2014gia,Faccioli:2014cqa}.

The color-octet LDMEs that are obtained from the fit are
\begin{subequations}
\begin{eqnarray}
\langle {\cal O}^{\psi(2S)} ({}^3S_1^{[8]}) \rangle &=& (-1.57 \pm 2.80)
\times 10^{-3}
\textrm{~GeV}^3,
\\
\langle {\cal O}^{\psi(2S)} ({}^1S_0^{[8]}) \rangle &=& (+ 3.14 \pm 0.79)
\times 10^{-2}
\textrm{~GeV}^3,
\\
\frac{\langle {\cal O}^{\psi(2S)} ({}^3P_0^{[8]}) \rangle}{m_c^2}
&=& (-1.14 \pm 1.21)
\times 10^{-3}
\textrm{~GeV}^3.
\end{eqnarray}
\end{subequations}

The uncertainties that are shown above are correlated. The correlation
matrix of the uncertainties in $\langle {\cal O}^{\psi(2S)}
({}^3S_1^{[8]}) \rangle$, $\langle {\cal O}^{\psi(2S)} ({}^1S_0^{[8]})
\rangle$, and $\langle {\cal O}^{\psi(2S)} ({}^3P_0^{[8]})
\rangle/m_c^2$, respectively, is
\begin{equation}
C^{\psi(2S)} =
\begin{pmatrix}
\phantom{-}7.85& -14.7& \phantom{-}3.36 \\
-14.7& \phantom{-}62.2& -5.52 \\
\phantom{-}3.36& -5.52& \phantom{-}1.46
\end{pmatrix} \times 10^{-6}\textrm{~GeV}^6.
\end{equation}

It is useful to examine the correlation matrix of relative uncertainties,
$\bar C^{\psi(2S)}$, whose components are defined by
\begin{equation}
\bar C^{\psi(2S)}_{nm} =
\frac{C^{\psi(2S)}_{nm}}{{\cal O}_n {\cal O}_m},
\end{equation}
where ${\cal O}_n$ is the central value of the $n$th LDME. Then $\bar
C^{\psi(2S)}$ is given by
\begin{equation}
\bar C^{\psi(2S)} =
\begin{pmatrix}
32.0 & 2.98  & 18.9 \\
2.98 & 0.63 & 1.55 \\
18.9 & 1.55  & 11.3
\end{pmatrix} \times 10^{-1}.
\end{equation}
The normalized eigenvectors of $\bar C^{\psi(2S)}$ are
\begin{equation}
v_1^{\psi(2S)} =
\begin{pmatrix}
0.858\\ 0.0780\\ 0.508
\end{pmatrix},
\quad
v_2^{\psi(2S)} =
\begin{pmatrix}
\phantom{-}0.181\\ \phantom{-}0.879\\ -0.441
\end{pmatrix},
\quad
v_3^{\psi(2S)} =
\begin{pmatrix}
-0.481\\ \phantom{-}0.470\\ \phantom{-}0.740
\end{pmatrix},
\end{equation}
and the corresponding eigenvalues are
$\lambda_1^{\psi(2S)} = 4.34$,
$\lambda_2^{\psi(2S)} = 4.67 \times 10^{-2}$,
and
$\lambda_3^{\psi(2S)} = 1.96 \times 10^{-3}$.
The eigenvector $v_2^{\psi(2S)}$ is predominantly
${}^1S_0^{[8]}$ and its uncertainty [$(\lambda_2^{\psi(2S)})^{1/2}$] is
fairly small. On the other hand, the eigenvector
$v_1^{\psi(2S)}$ has a very large uncertainty
[$(\lambda_1^{\psi(2S)})^{1/2}$]. Hence, the ${}^3S_1^{[8]}$ and
${}^3P_0^{[8]}$ LDMEs can vary together in a correlated way that tends
to preserve the ${}^1S_0^{[8]}$ dominance. (Recall that the SDCs for
these channels have opposite signs.)  The eigenvector
$v_3^{\psi(2S)}$ has a very small uncertainty
[$(\lambda_3^{\psi(2S)})^{1/2}$] and, therefore, the anticorrelated
variation of the ${}^3S_1^{[8]}$ and ${}^3P_0^{[8]}$ LDMEs is highly
constrained.

\subsection{Production of $\bm{\chi_{c1}}$ and $\bm{\chi_{c2}}$}

\begin{figure}
\epsfig{file=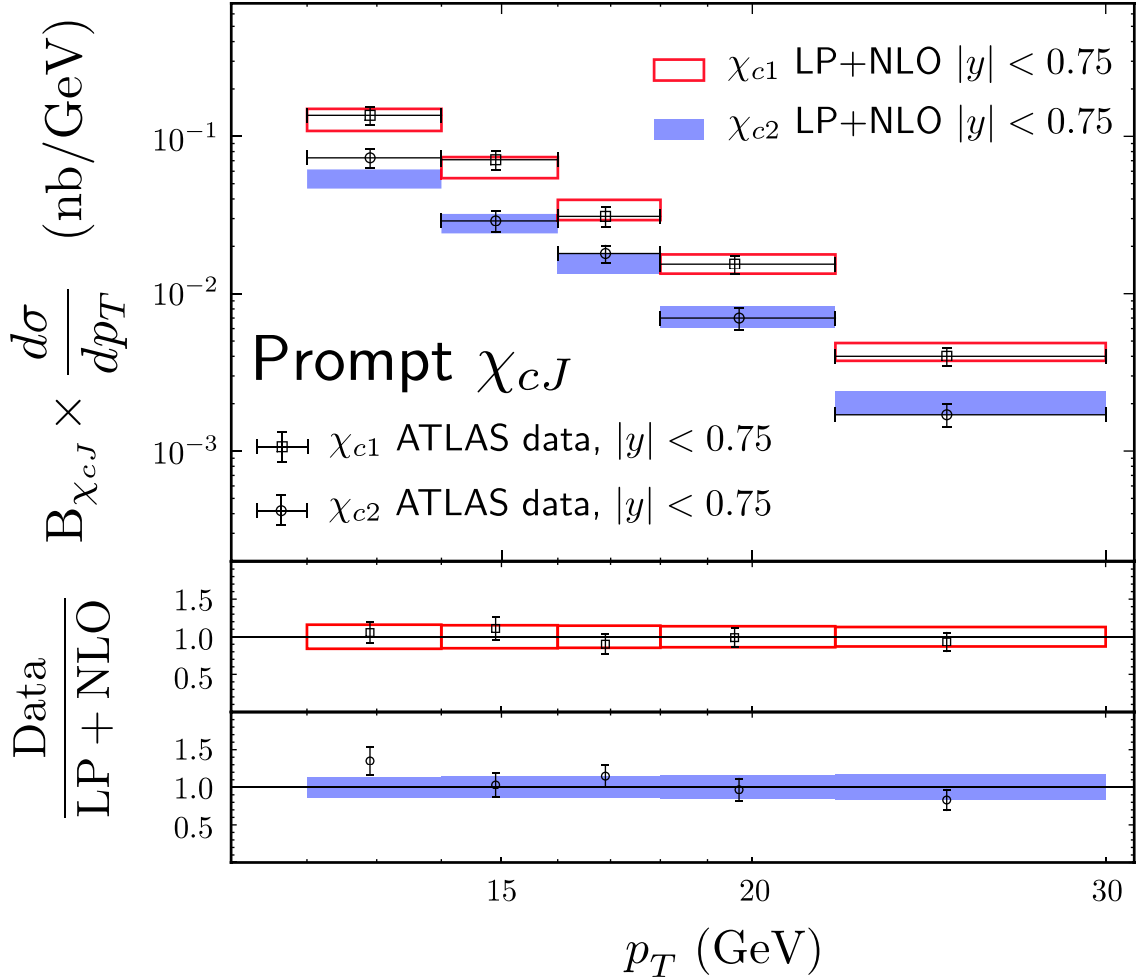,width=15cm}
\caption{\label{fig:chi_fit}%
The differential cross sections for prompt $\chi_{c1}$ and
$\chi_{c2}$ production at the LHC ($\sqrt{s}=7$~TeV).
$B_{\chi_{cJ}} =
{\rm Br} [\chi_{cJ} \to J/\psi + \gamma] \times
{\rm Br} [J/\psi \to \mu^+ \mu^-]$.
}
\end{figure}

We determine the two LDMEs for $\chi_{cJ}$ by fitting to ATLAS cross-section
data~\cite{ATLAS:2014ala}. In order to suppress possible nonfactorizing
contributions we fit only to data for which $p_T$ is greater than
$11$~GeV. We ignore feeddown contributions. The $\psi(2S)$ decays into
$\chi_{c1} \gamma$ and $\chi_{c2} \gamma$ with branching ratios of
$9.55\%$ and $9.11\%$, respectively. These contributions amount to only
a few percent of the measured cross sections and are much smaller than
the theoretical uncertainties.

\begin{figure}
\epsfig{file=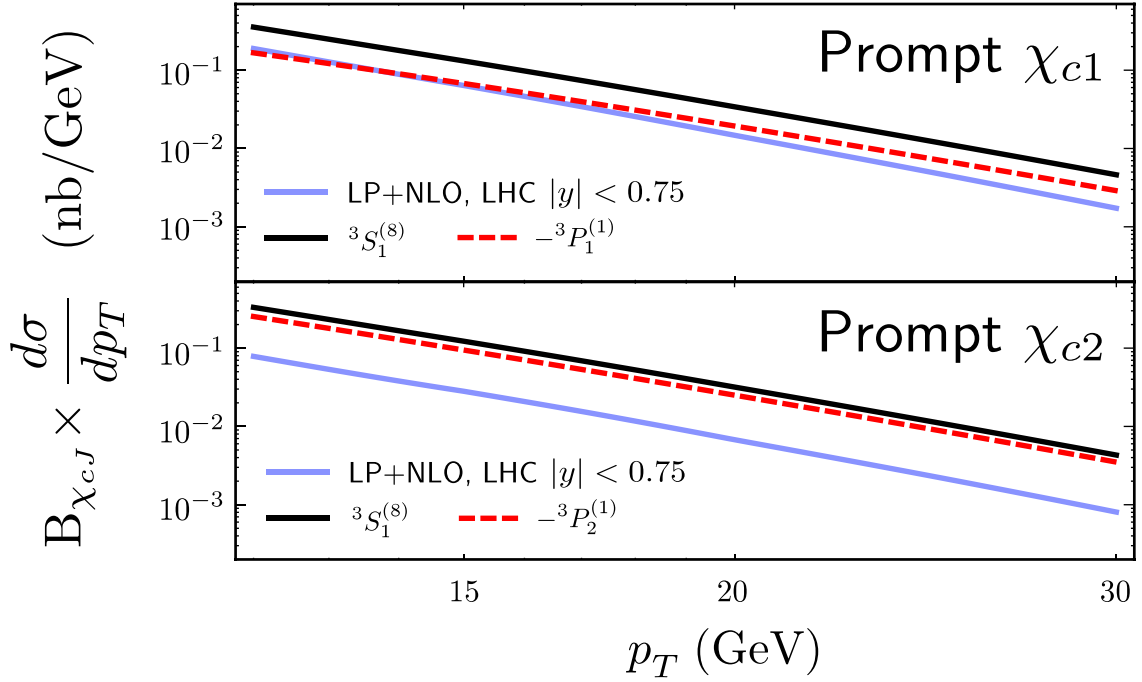,width=15cm}
\caption{\label{fig:chi_comp}%
Contributions of the individual channels to the
differential cross sections for prompt $\chi_{c1}$ and
$\chi_{c2}$ production at the LHC ($\sqrt{s}=7$~TeV).
$B_{\chi_cJ} =
{\rm Br} [\chi_{cJ} \to J/\psi + \gamma] \times
{\rm Br} [J/\psi \to \mu^+ \mu^-]$.
}
\end{figure}

The fitted LP$+$NLO $\chi_{c1}$ and $\chi_{c2}$ cross sections are
compared with the data in Fig.~\ref{fig:chi_fit}. 
We do not consider the $\chi_{c0}$ cross section because the
$\chi_{c0}$ branching ratio to $J/\psi \gamma$ is small and the corresponding
contribution to the prompt $J/\psi$ cross section is negligible.
Again, we obtain a good fit to data, with
$\chi^2/{\rm d.o.f.} = 1.19/8$. The contributions of the individual
channels to the prompt-$\chi_{c1}$ and prompt-$\chi_{c2}$ cross
sections are shown in Fig.~\ref{fig:chi_comp}. There are substantial
cancellations between the contributions of the ${}^3S_1^{[8]}$ and
${}^3P_J^{[1]}$ channels.

The resulting LDMEs are
\begin{subequations}
\begin{eqnarray}
\langle {\cal O}^{\chi_c} ({}^3S_1^{[8]}) \rangle &=&
(5.74 \pm 1.31) \times 10^{-3}
\textrm{~GeV}^3,
\\
\frac{\langle {\cal O}^{\chi_c} ({}^3P_0^{[1]}) \rangle}{m_c^2} &=&
(3.53 \pm 1.08) \times 10^{-2}
\textrm{~GeV}^3.
\end{eqnarray}
\end{subequations}
The correlation matrix of the uncertainties in $\langle {\cal
O}^{\chi_c} ({}^3S_1^{[8]}) \rangle$ and $\langle {\cal O}^{\chi_c}
({}^3P_0^{[1]}) \rangle/m_c^2$, respectively, is
\begin{equation}
C^{\chi_c} =
\begin{pmatrix}
1.71 & ~14.0 \\
14.0 & ~117
\end{pmatrix}
\times 10^{-6}\textrm{ GeV}^6.
\end{equation}
The relative uncertainties in these LDMEs are fairly small, but there
are substantial correlations between them.
The correlation matrix of relative uncertainties is
\begin{equation}
\bar C^{\chi_c} =
\begin{pmatrix}
5.18 & ~6.91 \\
6.91 & ~9.39
\end{pmatrix} \times 10^{-2}.
\end{equation}
The normalized eigenvectors of $\bar C^{\chi_c}$ are
\begin{equation}
{v}_1^{\chi_c} =
\begin{pmatrix}
0.595 \\ 0.804
\end{pmatrix},
\quad
{v}_2^{\chi_c} =
\begin{pmatrix}
\phantom{-}0.804 \\ -0.595
\end{pmatrix},
\end{equation}
and the corresponding eigenvalues are $\lambda_1^{\chi_c} = 0.145$ and
$\lambda_2^{\chi_c} = 5.70 \times 10^{-4}$. We see
that, while the eigenvector $v_1^{\chi_c}$ has a small uncertainty, the
${}^3S_1^{[8]}$ and ${}^3P_0^{[1]}$ LDMEs can vary in a correlated way.
However, from the very small uncertainty of the eigenvector
$v_2^{\chi_c}$, we see that anticorrelated variation of these LDMEs is
highly constrained.

At leading order in $v$,
the color-singlet LDME is related to the derivative of the wave function
at the origin as
\begin{equation}
\langle {\cal O} ({}^3P_0^{[1]}) \rangle_{\chi_c}
= 2 N_c\, \frac{3}{4 \pi} |R'(0)|^2.
\end{equation}
The value of the color-singlet LDME that we obtained from our fit
corresponds to $|R'(0)|^2 = 0.055 \pm 0.017$~GeV${}^5$. This is
consistent with the value $|R'(0)|^2 = 0.075$~GeV${}^5$ that was
obtained in Ref.~\cite{Eichten:1995ch} by using the Buchm\"uller-Tye
potential. It is also consistent with the value that was determined in
Ref.~\cite{Chung:2008km} from the two-photon decay rates of the
$\chi_{c0}$ and the $\chi_{c2}$, namely, $\langle {\cal O}
({}^3P_0^{[1]}) \rangle_{\chi_c}=0.060^{+0.043}_{-0.029}$~GeV${}^5$,
which corresponds to $|R'(0)|^2 =
0.042^{+0.030}_{-0.020}$~GeV${}^5$.

\subsection{Production of prompt $\bm{J/\psi}$}

We determine the $J/\psi$ LDMEs by fitting to the
CDF~\cite{Acosta:2004yw} and CMS~\cite{Chatrchyan:2011kc, Khachatryan:2015rra} 
prompt-$J/\psi$ cross-section data. In order to suppress
possible nonfactorizing contributions, we fit only to data for $p_T$
greater than $10$ GeV. We compute the feeddown contributions from the
decays of $\psi(2S)$, $\chi_{c1}$, and $\chi_{c2}$ by making use of the
LDMEs that were determined in the preceding sections. The prompt-$J/\psi$ 
cross section is given by
\begin{eqnarray}
\frac{d \sigma_{J/\psi}^{\rm prompt} }{d p_T}
&=&
\frac{d \sigma_{J/\psi}^{\rm direct} }{d p_T}
+
\frac{d \sigma_{\psi(2S)} }{d p_T^{\psi(2S)}}
{\rm Br}[\psi(2S) \to J/\psi + X]
+
\frac{d \sigma_{\chi_{c1}} }{d p_T^{\chi_{c1}}}
{\rm Br}[\chi_{c1} \to J/\psi + \gamma]\nonumber\\
&&+
\frac{d \sigma_{\chi_{c2}} }{d p_T^{\chi_{c2}}}
{\rm Br}[\chi_{c2} \to J/\psi + \gamma].
\end{eqnarray}
Here, we ignore the feeddown contribution from the decay of the
$\chi_{c0}$. As we have mentioned, the $\chi_{c0}$ decays into $J/\psi
\gamma$ with a small branching ratio, and the contribution to the prompt
$J/\psi$ cross section is negligible. $p_T$ is the transverse momentum
of the $J/\psi$, and $p_T^{H}$ is the transverse momentum of
$H=\psi(2S)$, $\chi_{cJ}$.
In the feeddown contributions, we take $p_T^H$ to be
\begin{equation}
p_T^H = \frac{m_H}{m_{J/\psi}} p_T.
\label{feeddown-momenta}
\end{equation}
The relation
(\ref{feeddown-momenta}) is derived by neglecting the 3-momentum of the
$J/\psi$ in the $H$ rest frame in comparison with
$m_{J/\psi}$.\footnote{We have estimated the effects of corrections to
this relation on the contributions of $\chi_{c1}$ and $\chi_{c2}$
feeddown to the $J/\psi$ unpolarized and polarized cross sections. In
these estimates, we computed the angular distribution of the $J/\psi$
momentum in the $\chi_{cJ}$ rest frame by making use of the formalism of
Ref.~\cite{Faccioli:2011be}, and we included the E1, M2, and E3
electromagnetic transition amplitudes, taking the M2 and E3 amplitudes
to be given by the central values of the measurement of the CLEO
Collaboration \cite{Artuso:2009aa}. We find that the corrections to the
feeddown contributions are no more than $8\%$ in any of the $J/\psi$
polarization channels. Furthermore, the corrections are essentially
flat as functions of $p_T$, deviating by only about $1\%$ over the range
$10~{\rm GeV}\leq p_T\leq 100~{\rm GeV}$, with almost all of the
deviation occurring between $10$~GeV and $15$~GeV. Hence, the
corrections have little effect on the shapes of the cross sections and
can be absorbed into normalization shifts of the LDMEs of a few
percent or less.}

We take the value of the color-singlet LDME that has been obtained
from the electromagnetic decay rate \cite{Bodwin:2007fz}: $\langle {\cal
O}^{J/\psi} ({}^3S_1^{[1]}) \rangle = 1.32$~GeV${}^3$. Again, the
contribution from the color-singlet channel is much smaller than the
theoretical uncertainties, ranging from $4\%$ for the bin
$10\textrm{~GeV}<p_T<11\textrm{~GeV}$ to $0.2\%$ for the bin
$95\textrm{~GeV}<p_T<120\textrm{~GeV}$ in comparison with the
direct $J/\psi$ cross section.

\begin{figure}
\epsfig{file=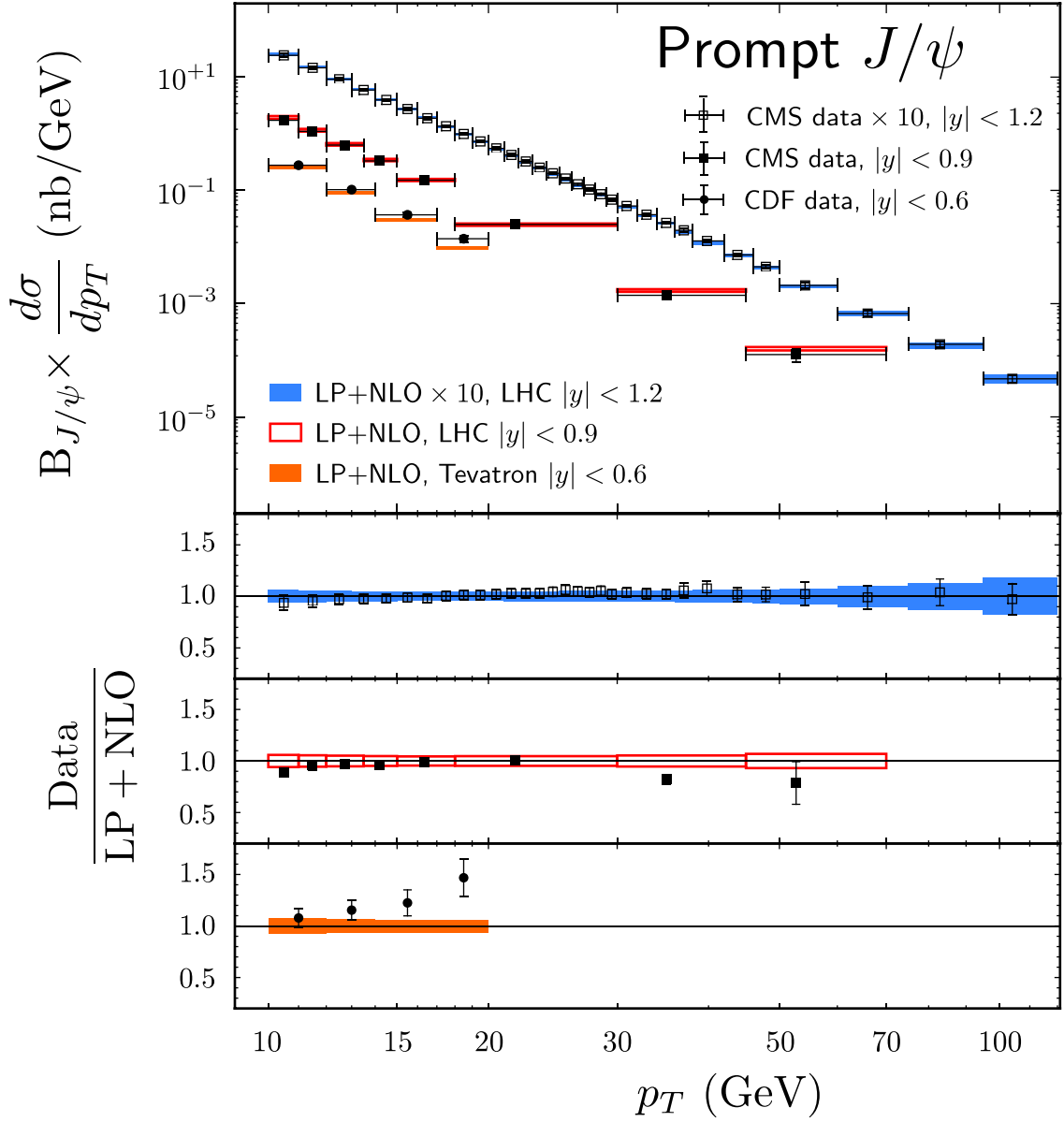,width=15cm}
\caption{\label{fig:jpsi_fit}%
The differential cross section for prompt $J/\psi$
production at the Tevatron ($\sqrt{s}=1.96$~TeV) and the LHC
($\sqrt{s}=7$~TeV).
$B_{J/\psi} = {\rm Br} [J/\psi \to \mu^+ \mu^-]$.
}
\end{figure}

\begin{figure}
\epsfig{file=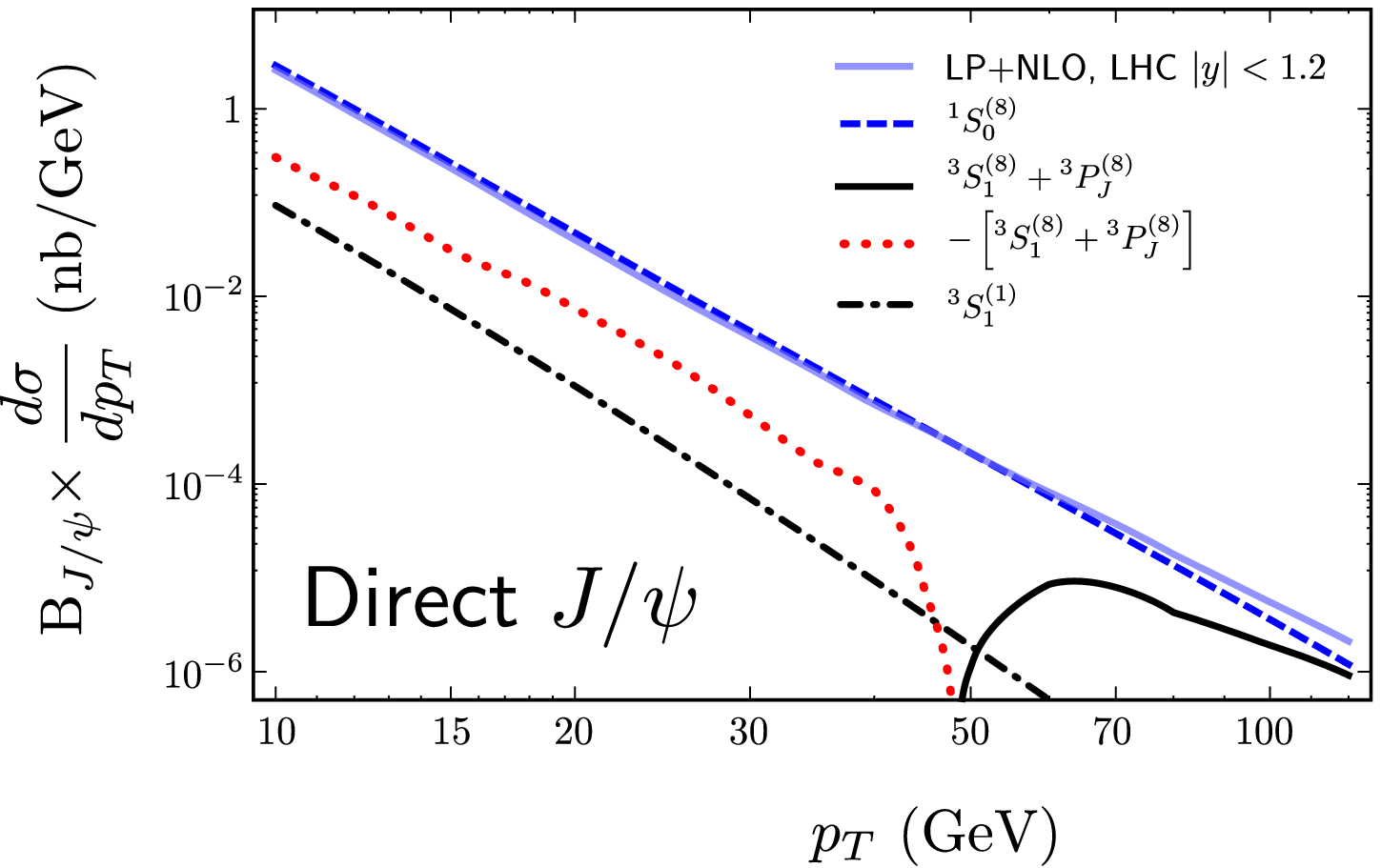,width=15cm}
\caption{\label{fig:jpsi_comp}%
Contributions of the individual channels to the differential
cross section for direct $J/\psi$ production at the LHC
($\sqrt{s}=7$~TeV).
$B_{J/\psi} = {\rm Br} [J/\psi \to \mu^+ \mu^-]$.
}
\end{figure}

We obtain a good fit to the data, with $\chi^2/{\rm d.o.f.} = 8.20/40$. The
fitted LP$+$NLO cross section is shown in comparison with the data in
Fig.~\ref{fig:jpsi_fit}. The contributions of the individual channels to
the direct $J/\psi$ cross section are shown in Fig.~\ref{fig:jpsi_comp}.
The direct $J/\psi$ cross section is dominated by the ${}^1S_0^{[8]}$
channel at all values of $p_T$ between $10$~GeV and $100$~GeV.

The color-octet LDMEs that are obtained from the fit are
\begin{subequations}
\label{Jpsi-LDMEs}
\begin{eqnarray}
\langle {\cal O}^{J/\psi} ({}^3S_1^{[8]}) \rangle &=&
(-7.13 \pm 3.64)
\times 10^{-3}
\textrm{~GeV}^3,
\\
\langle {\cal O}^{J/\psi} ({}^1S_0^{[8]}) \rangle &=&
(+1.10 \pm 0.14)
\times 10^{-1}
\textrm{~GeV}^3,
\\
\frac{\langle {\cal O}^{J/\psi} ({}^3P_0^{[8]}) \rangle}{m_c^2}
&=&
(-3.12 \pm 1.51)
\times 10^{-3}
\textrm{~GeV}^3.
\end{eqnarray}
\end{subequations}
The correlation matrix of the uncertainties in
$\langle {\cal O}^{J/\psi} ({}^3S_1^{[8]}) \rangle$,
$\langle {\cal O}^{J/\psi} ({}^1S_0^{[8]}) \rangle$, and
$\langle {\cal O}^{J/\psi} ({}^3P_0^{[8]}) \rangle/m_c^2$, respectively,
is
\begin{equation}
C^{J/\psi} =
\begin{pmatrix}
\phantom{-}13.3 & -38.2 & \phantom{-}5.48 \\
-38.2 & \phantom{-}188 & -14.6 \\
\phantom{-}5.48 & -14.6 & \phantom{-}2.29
\end{pmatrix} \times 10^{-6}\textrm{~GeV}^6.
\end{equation}

The correlation matrix of relative uncertainties is
\begin{equation}
\bar C^{J/\psi} =
\begin{pmatrix}
26.1 & 4.88 & 24.7 \\
4.88 & 1.55 & 4.26 \\
24.7 & 4.26 & 23.5
\end{pmatrix} \times 10^{-2}.
\end{equation}
The normalized eigenvectors of $\bar C^{J/\psi}$ are
\begin{equation}
v_1^{J/\psi} =
\begin{pmatrix}
0.719 \\ 0.131 \\ 0.682
\end{pmatrix},
\quad
v_2^{J/\psi} =
\begin{pmatrix}
\phantom{-}0.168 \\ \phantom{-}0.920 \\ -0.354
\end{pmatrix},
\quad
v_3^{J/\psi} =
\begin{pmatrix}
-0.674 \\ \phantom{-}0.369 \\ \phantom{-}0.640
\end{pmatrix},
\end{equation}
and the corresponding eigenvalues are
$\lambda_1^{J/\psi} = 0.504$,
$\lambda_2^{J/\psi} = 8.06 \times 10^{-3}$,
and
$\lambda_3^{J/\psi} = 2.35 \times 10^{-4}$.
As is the case for the $\psi(2S)$, the eigenvector that is predominantly
${}^1S_0^{[8]}$, namely, $v_2^{J/\psi}$, has a fairly small
uncertainty. However, the eigenvector $v_1^{J/\psi}$ has a very large
uncertainty. Therefore, variations of the ${}^3S_1^{[8]}$ and
${}^3P_0^{[8]}$ LDMEs are correlated and tend to preserve the
${}^1S_0^{[8]}$ dominance. (Recall that the SDCs for these channels have
opposite signs.) The very small uncertainty of the eigenvector
$v_3^{J/\psi}$ means that the anticorrelated variation of the
${}^3S_1^{[8]}$ and ${}^3P_0^{[8]}$ LDMEs is highly constrained.

\begin{figure}
\epsfig{file=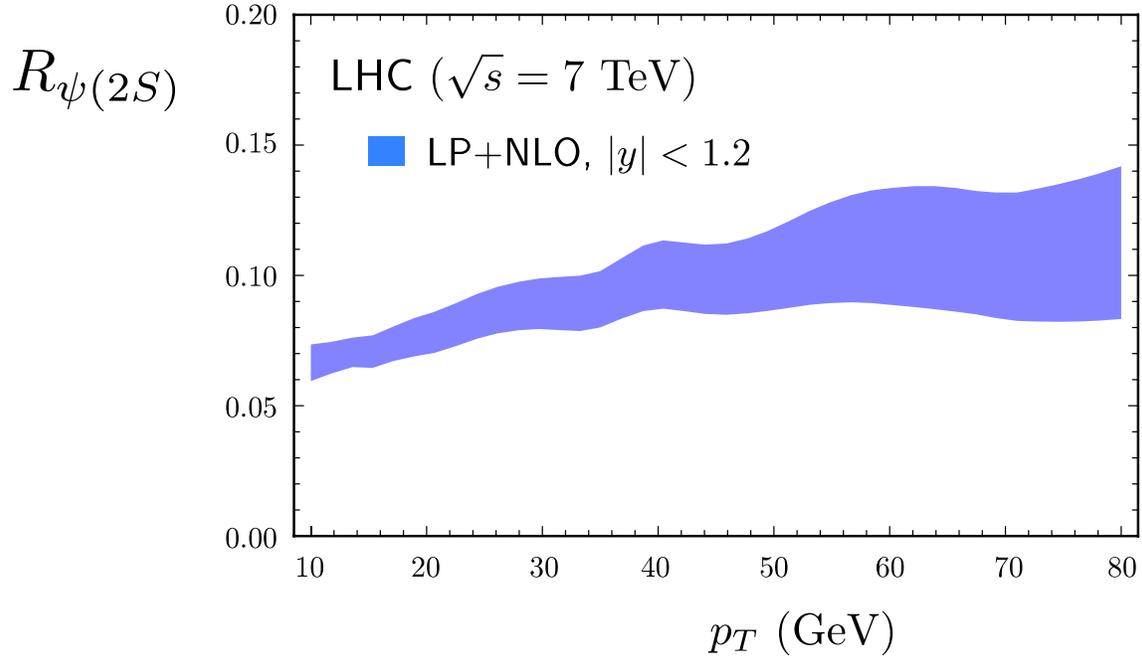,width=15cm}
\caption{\label{fig:psi2s_ratio}%
Fraction of prompt $J/\psi$'s produced in feeddown from $\psi(2S)$ decays
at the LHC ($\sqrt{s}=7$~TeV).
}
\end{figure}

\begin{figure}
\epsfig{file=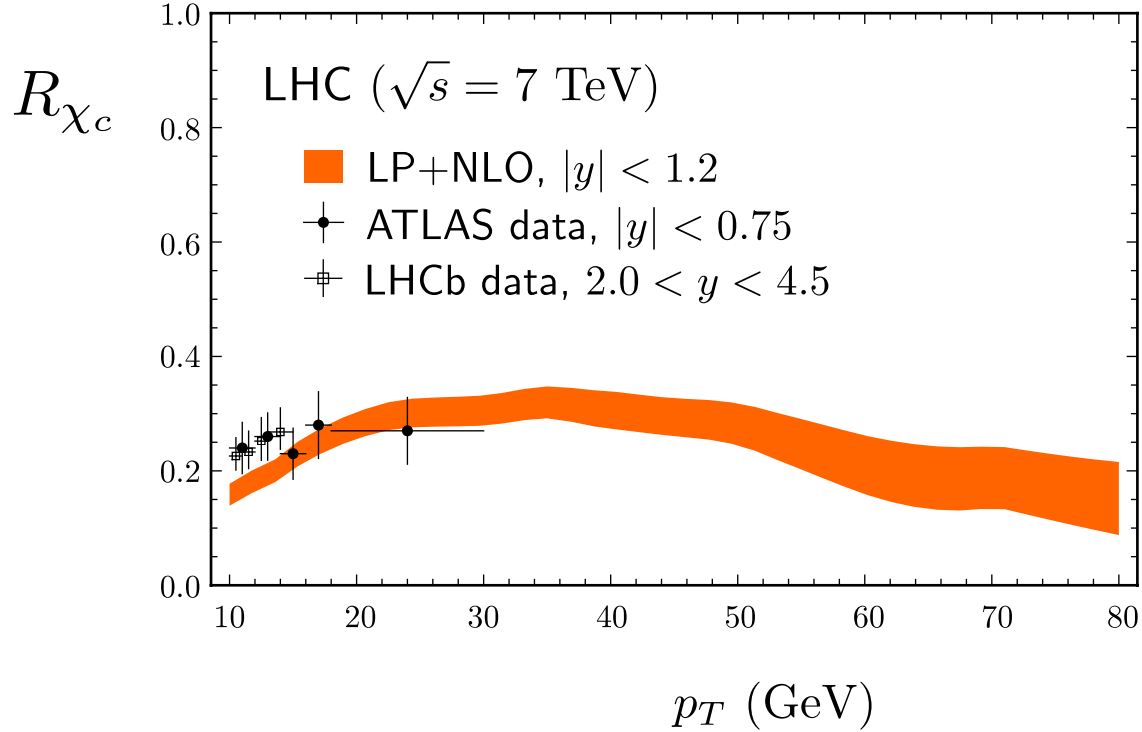,width=15cm}
\caption{\label{fig:chi_ratio}%
Fraction of prompt $J/\psi$'s produced in feeddown from $\chi_{c1}$
and $\chi_{c2}$ decays at the LHC ($\sqrt{s}=7$~TeV).
}
\end{figure}

\section{Predictions from Extracted LDME\lowercase{s}
\label{sec:predictions}}

In this section, we use the LDMEs that we have extracted from the fits
to cross sections to make predictions of cross-section ratios and
polarizations. We estimate the uncertainties in these predictions by
making use of the eigenvectors and eigenvalues of the LDME uncertainty
correlation matrices.  In the expression for each prediction, we write
the LDMEs in terms of the eigenvectors. Then, we vary each eigenvector
about its central value by an amount that is equal to the square root of
its eigenvalue. We take the resulting variation in the prediction as the
uncertainty in the prediction from variations of that eigenvector.
Finally, we estimate the total uncertainty in the prediction by adding the
uncertainties from the variations of the individual eigenvectors in
quadrature.

\subsection{Ratios $\bm{R_H}$}

We can use our predictions for the $J/\psi$, $\psi(2S)$, and
$\chi_{cJ}$ cross sections and the LDMEs that we have extracted to
compute the ratios $R_H$, which are defined by
\begin{equation}
R_H =
\frac{
{\rm Br}[H \to J/\psi+X] \times
d \sigma_H/d p_T^H}{d \sigma_{J/\psi}^{\rm prompt} /d p_T},
\end{equation}
where $p_T^H$ is given in Eq.~(\ref{feeddown-momenta}). In
Figs.~\ref{fig:psi2s_ratio} and \ref{fig:chi_ratio} we show our results
for $R_{\psi(2S)}$ and $R_{\chi_c}\equiv R_{\chi_{c1}}+R_{\chi_{c2}}$, 
respectively. As can be seen from
Fig.~\ref{fig:chi_ratio}, our prediction for $R_{\chi_c}$ lies
systematically below the ATLAS~\cite{ATLAS:2014ala} and
LHCb~\cite{LHCb:2012af} measurements for $p_T<15$~GeV. This discrepancy
occurs because the prediction for the numerator of $R_{\chi_c}$ lies
slightly below the data at low $p_T$, while the prediction for the
denominator of $R_{\chi_c}$ lies slightly above the data at low $p_T$.
However, the predictions for both the numerator and the denominator
agree with the data within uncertainties. We also note that
corrections to the relation (\ref{feeddown-momenta}) for the $J/\psi$
momentum would increase the theoretical prediction for $R_{\chi_c}$ by a
few percent.

\subsection{Polarization predictions}
We now compute prompt-$\psi(2S)$ and prompt-$J/\psi$
polarizations by making use of the LDMEs that we have determined from
fits to the cross-section data. We also compute the effects of feeddown
from the $\psi(2S)$ and $\chi_{cJ}$ states on the polarizations of the
prompt $J/\psi$'s.

For $J=1$ states, one measure of
the polarization is the polarization parameter
$\lambda_\theta$, which is defined as
\begin{equation}
\lambda_\theta =
\frac{\sigma - 3 \sigma_L}{\sigma+ \sigma_L},
\end{equation}
where $\sigma$ and $\sigma_L$ are the polarization-summed and
longitudinal cross sections, respectively. If the $J=1$ state is completely
transversely (longitudinally) polarized, then $\sigma_L = 0$ ($\sigma_L
= \sigma$), and $\lambda_\theta=+1$ ($\lambda_\theta=-1$). If the
$J=1$ state is unpolarized, then $\sigma = 3 \sigma_L$, and
$\lambda_\theta =0$.

\begin{figure}
\epsfig{file=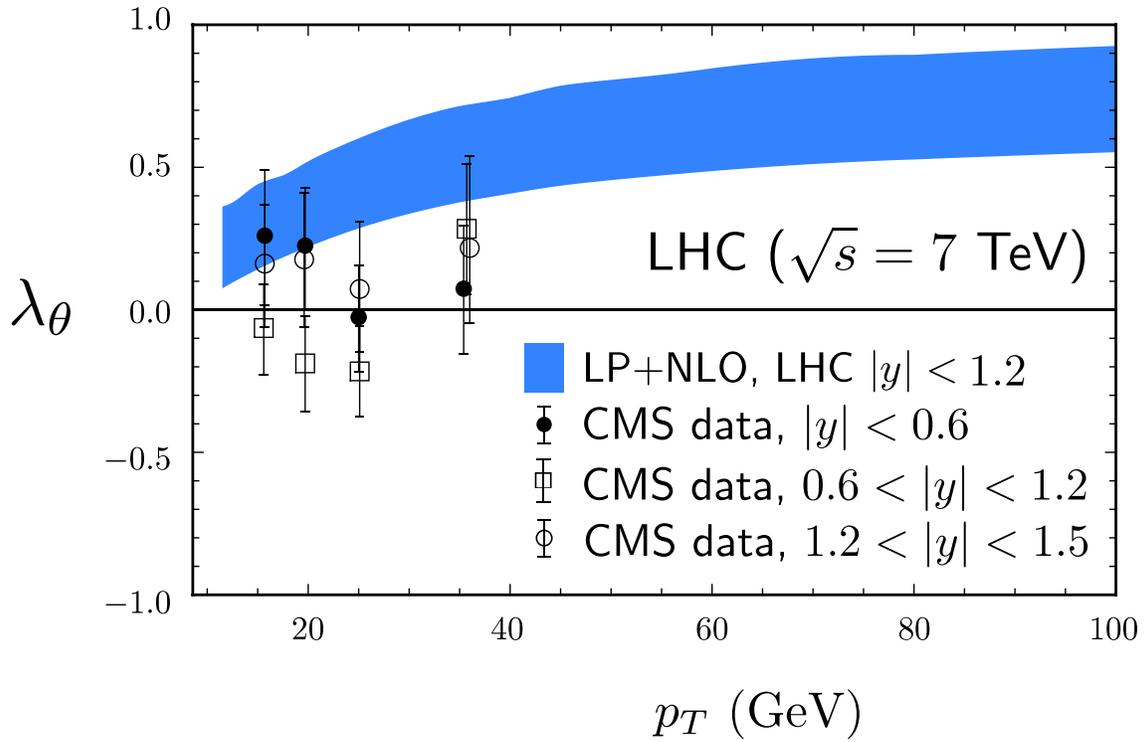,width=15cm}
\caption{\label{fig:psi2s_pol_cms}%
Polarization of prompt $\psi(2S)$ at the LHC ($\sqrt{s}=7$~TeV).
}
\end{figure}

\begin{figure}
\epsfig{file=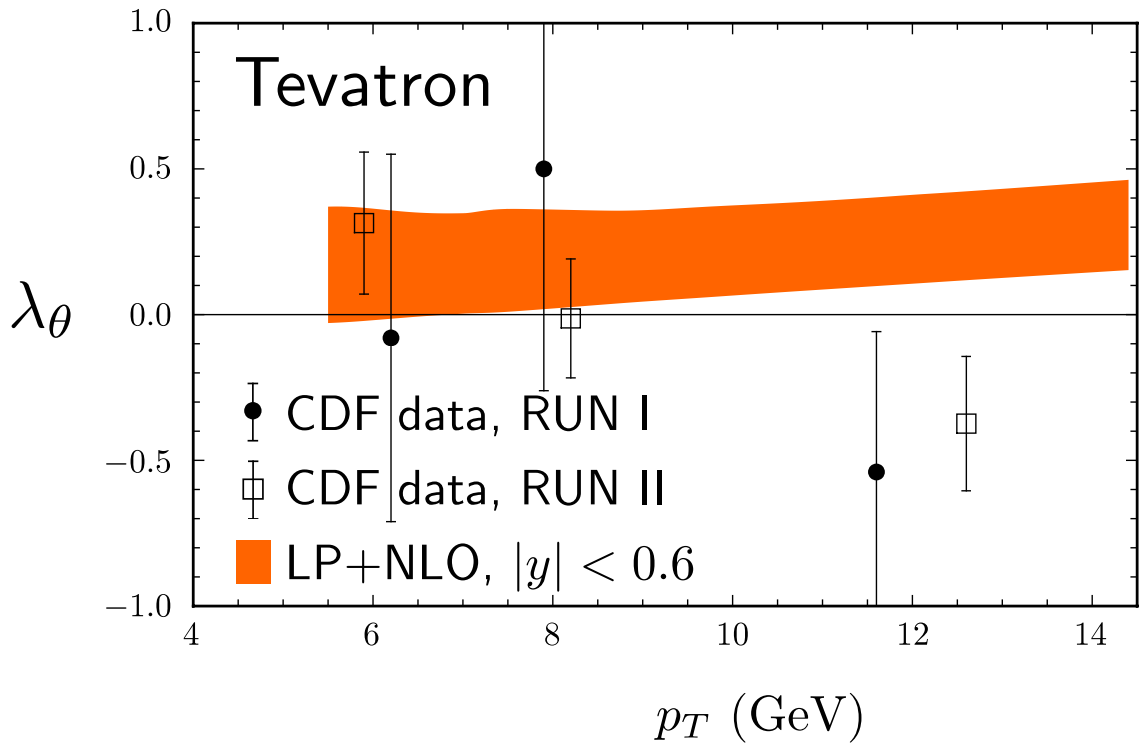,width=15cm}
\caption{\label{fig:psi2s_pol_cdf}%
Polarization of prompt $\psi(2S)$ at the Tevatron ($\sqrt{s}=1.96$~TeV).
}
\end{figure}

We show the polarization of the $\psi(2S)$ as produced at the LHC at
$\sqrt{s}=7$~TeV and at the Tevatron at $\sqrt{s}=1.96$~TeV in
Figs.~\ref{fig:psi2s_pol_cms} and \ref{fig:psi2s_pol_cdf}, respectively.
The prediction for the CMS polarization is in fair
agreement with the CMS data~\cite{Chatrchyan:2013cla}.
The prediction for the polarization at the Tevatron is in rough
agreement with the CDF Run~I~\cite{Affolder:2000nn} and
Run~II~\cite{Abulencia:2007us} data, given the very large error
bars. (Although the CDF Run~I data were taken at $\sqrt{s}=1.8$~TeV,
rather than at $\sqrt{s}=1.96$~TeV, this energy shift produces a
negligible change in the polarization prediction.) The
predicted $\psi(2S)$ polarization grows as $p_T$ increases, owing to the
fact that the ${}^1S_0^{[8]}$ channel is no longer dominant at large
$p_T$. Hence, measurements of the $\psi(2S)$ polarization at larger
values of $p_T$ would provide an important test of the theoretical
prediction.

The longitudinal prompt-$J/\psi$ cross section, including the feeddown
contributions from the decays of the $\psi(2S)$, the $\chi_{c1}$, and the
$\chi_{c2}$ is computed as follows:
\begin{eqnarray}
\frac{d \sigma_{J/\psi (\lambda = 0)}^{\rm prompt} }{d p_T}
&=&
\frac{d \sigma_{J/\psi (\lambda = 0)}^{\rm direct} }{d p_T}
+
\frac{d \sigma_{\psi(2S) (\lambda = 0)} }{d p_T^{\psi(2S)}}
{\rm Br}[\psi(2S) \to J/\psi + X]
\nonumber \\
&& +
\frac{1}{2}
\bigg( \frac{d \sigma_{\chi_{c1} (\lambda = +1)} }{d p_T^{\chi_{c1}}}
+ \frac{d \sigma_{\chi_{c1} (\lambda = -1)} }{d p_T^{\chi_{c1}}} \bigg)
{\rm Br}[\chi_{c1} \to J/\psi + \gamma]
\nonumber \\
&& +
\bigg[
\frac{2}{3}
\frac{d \sigma_{\chi_{c2} (\lambda = 0)} }{d p_T^{\chi_{c2}}}
+
\frac{1}{2}
\bigg( \frac{d \sigma_{\chi_{c2} (\lambda = +1)} }{d p_T^{\chi_{c2}}}
+ \frac{d \sigma_{\chi_{c2} (\lambda = -1)} }{d p_T^{\chi_{c2}}} \bigg)
\bigg]
{\rm Br}[\chi_{c2} \to J/\psi + \gamma],
\label{prompt-psi-pol}
\end{eqnarray}
where $p_T^H$ is given by Eq.~(\ref{feeddown-momenta}). In deriving
Eq.~(\ref{prompt-psi-pol}), we have assumed that the polarization of the
$\psi(2S)$ is completely transferred to $J/\psi$ and that the decays
$\chi_{cJ} \to J/\psi + \gamma$ proceed through an E1 transition. In the
decays $\chi_{cJ} \to J/\psi + \gamma$, the higher multipole corrections
are poorly known, but they have little effect on the polarizations of
the $J/\psi$'s that are produced in $\chi_{cJ}$
decays~\cite{Faccioli:2011be}.

\begin{figure}
\epsfig{file=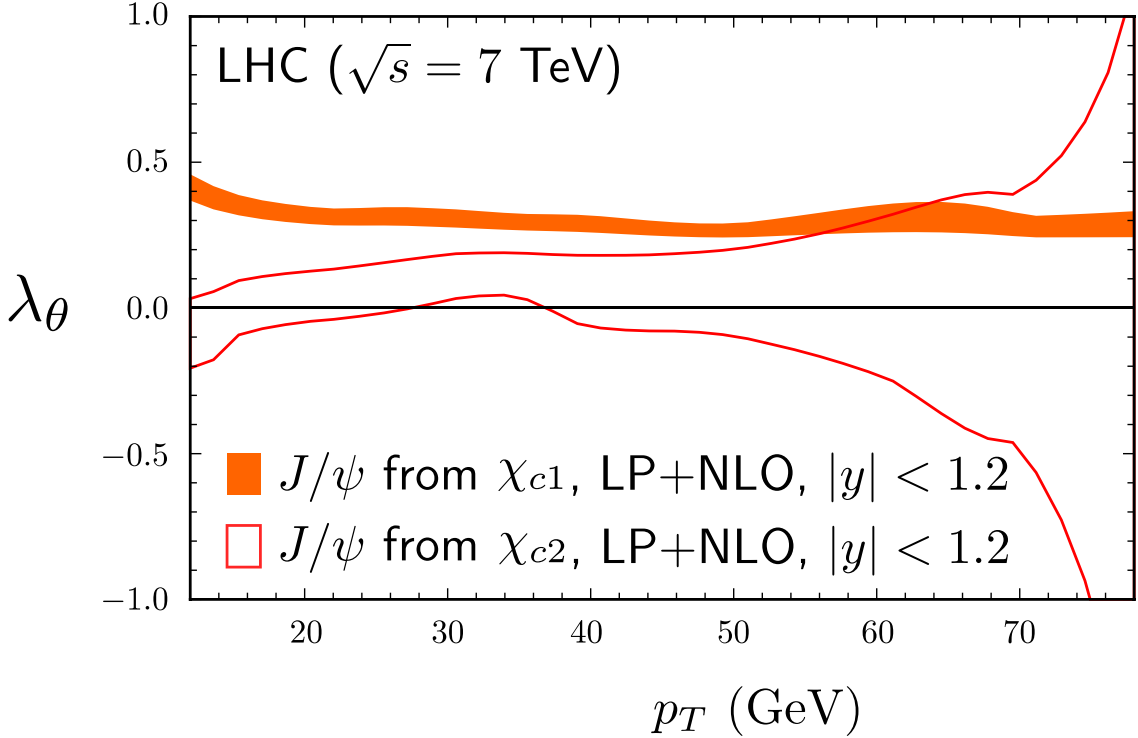,width=15cm}
\caption{\label{fig:chi_pol_cms}%
Polarization of $J/\psi$ from $\chi_{cJ}$ decays at the LHC
($\sqrt{s}=7$~TeV).
}
\end{figure}

\begin{figure}
\epsfig{file=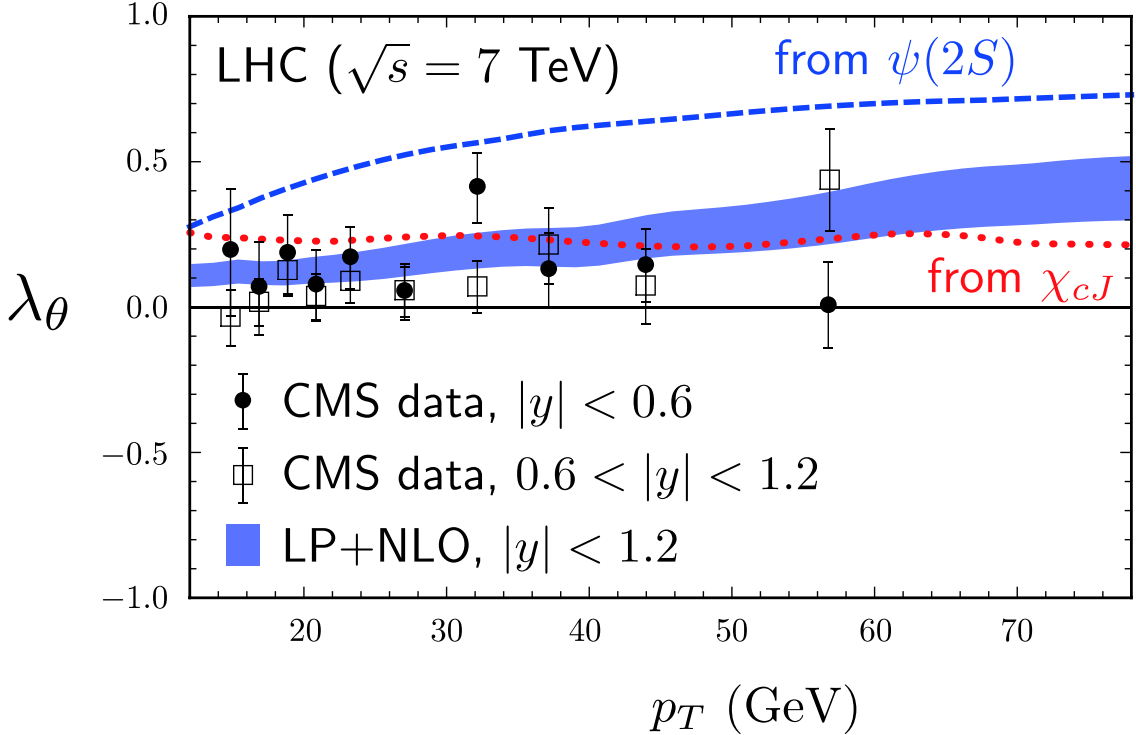,width=15cm}
\caption{\label{fig:jpsi_pol_cms}%
Polarization of prompt $J/\psi$'s at the LHC ($\sqrt{s}=7$~TeV). The
polarizations of $J/\psi$'s produced in feeddown from the $\psi(2S)$ and
$\chi_{cJ}$ states are shown with dashed and dotted lines, respectively.
}
\end{figure}

\begin{figure}
\epsfig{file=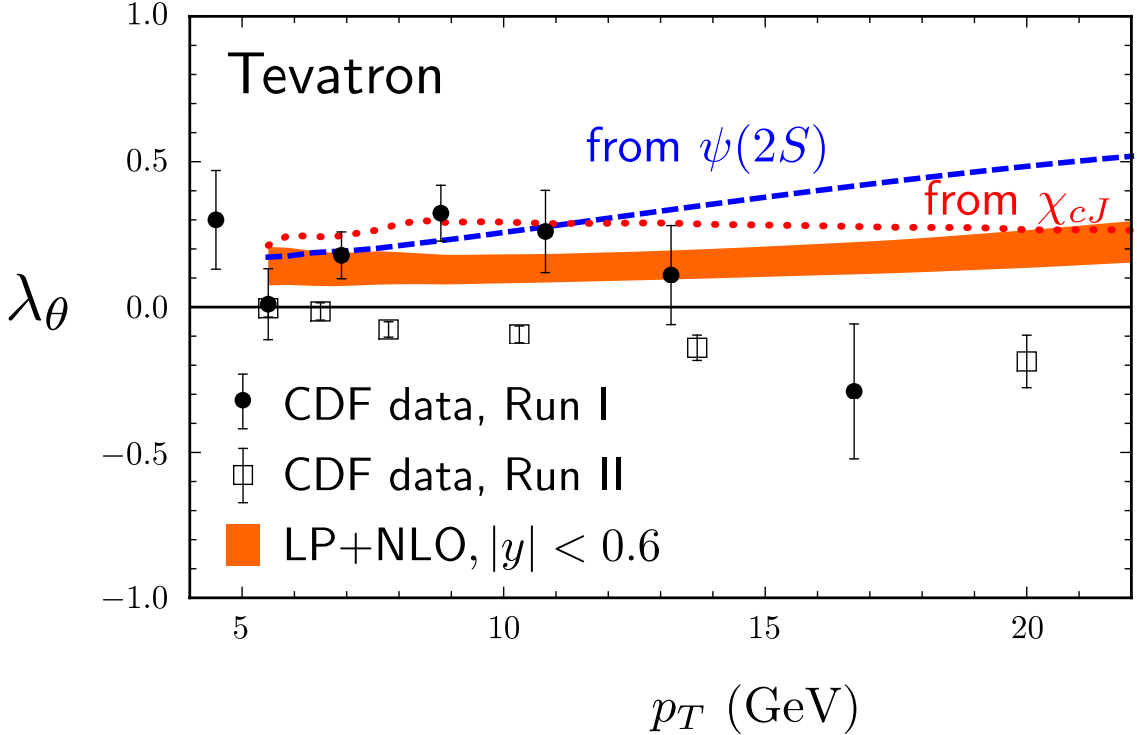,width=15cm}
\caption{\label{fig:jpsi_pol_cdf}%
Polarization of prompt $J/\psi$'s produced at the Tevatron 
($\sqrt{s}=1.96$~TeV). The
polarizations of $J/\psi$'s produced in feeddown from the $\psi(2S)$ and
$\chi_{cJ}$ states are shown with dashed and dotted lines, respectively.
}
\end{figure}

We show the polarization of $J/\psi$'s from
$\chi_{cJ}$ decays at the LHC at $\sqrt{s}=7$~TeV in
Fig.~\ref{fig:chi_pol_cms}. In Fig.~\ref{fig:jpsi_pol_cms}, we show the
polarization of prompt $J/\psi$'s produced at the LHC at
$\sqrt{s}=7$~TeV, including feeddown from the $\psi(2S)$ and the
$\chi_{cJ}$ states. The prediction is in good agreement with the 
CMS data~\cite{Chatrchyan:2013cla}. 
Finally, in Fig.~\ref{fig:jpsi_pol_cdf}, we show the polarization
of prompt $J/\psi$'s produced at the Tevatron at $\sqrt{s}=1.96$~TeV, 
including feeddown from
the $\psi(2S)$ and the $\chi_{cJ}$ states. The prediction is in good
agreement with the CDF Run~I data~\cite{Affolder:2000nn}, 
but disagrees with the CDF Run~II data~\cite{Abulencia:2007us}. 
(Although the CDF Run~I data were taken at $\sqrt{s}=1.8$~TeV, rather than at
$\sqrt{s}=1.96$~TeV, this energy shift produces a negligible change in
the polarization prediction.) We note that the predicted polarizations are almost the same for
the LHC and the Tevatron, while the CDF Run~II polarization data lies
significantly below the CMS polarization data.

The fairly small polarizations that are seen in the predictions for the
prompt $J/\psi$'s and $\psi(2S)$'s  are a consequence of the
dominance in the production rates of the ${}^1S_0^{[8]}$ channel, which,
of course, is completely unpolarized. This mechanism whereby small
polarizations can be obtained was noted previously in
Refs.~\cite{Chao:2012iv,Bodwin:2014gia,Faccioli:2014cqa}.

\section{Summary and Discussion \label{sec:summary}}

In this paper, we have computed, in the NRQCD factorization framework,
leading-power (LP) fragmentation corrections to production of the
charmonium states $J/\psi$, $\chi_{cJ}$, and $\psi(2S)$ in $p\bar p$
collisions at the Tevatron and in $pp$ collisions at the LHC.
Specifically, our calculation makes use of parton production cross
sections (PPCSs) through order $\alpha_s^3$ (NLO) and fragmentation
functions (FFs) through order $\alpha_s^2$. We have also
used the DGLAP equation to resum leading logarithms of $p_T^2/m_c^2$ to
all orders in $\alpha_s$. Our calculations take into account the effects
of feeddown from the $\psi(2S)$ and $\chi_{cJ}$ states on 
the prompt-$J/\psi$ cross sections and polarizations. 
Hence, the work in the
present paper is an extension and a refinement of the work in
Ref.~\cite{Bodwin:2014gia}, which also addressed LP corrections, but
which did not include computations of cross sections or polarizations
for the $\psi(2S)$ or $\chi_{cJ}$ states or include the effects of
feeddown from those states. We find that the LP corrections, beyond
those that are contained in fixed-order calculations through NLO in
$\alpha_s$, are substantial---typically of order $100\%$ at large $p_T$.
Owing to a partial cancellation between the LO and NLO contributions in
the ${}^3P_J^{[8]}$ channel, the LP corrections have a very significant
effect on the shape in that channel.

As was pointed out in Ref.~\cite{Bodwin:2014gia}, the all-orders
resummations of logarithms of $p_T^2/m_c^2$ have only small effects on
the predictions for the cross sections and polarizations. Hence, almost
all of the large additional LP corrections that we find arise from
nonlogarithmic contributions of order $\alpha_s^5$.

Our approach in calculating the LP fragmentation corrections is to use
the most accurate results for the PPCSs and FFs that are currently
available. This means that we have computed some, but not all, of the LP
contributions in order $\alpha_s^5$, {\it i.e.}, NNLO in terms of the
fixed-order calculations. In the case of gluon fragmentation, a complete
calculation of the
order-$\alpha_s^5$ contributions in the ${}^1S_0^{[8]}$ and
${}^3P_J^{[8]}$ channels would require the calculation of the NLO
corrections to the FFs for those channels. In the case of gluon
fragmentation, a complete calculation of the
order-$\alpha_s^5$ contributions in the ${}^3S_1^{[8]}$ channel would
require a calculation of the NNLO corrections to the FFs for that
channel and the NNLO corrections to the PPCSs. We expect these
uncalculated LP corrections in order $\alpha_s^5$ to be of comparable
size, in each channel, to the LP corrections that we have calculated in
this paper. Hence, the actual theoretical uncertainties may be much
larger than the estimates that we have obtained by varying the scales
$\mu_r$ and $\mu_f$. However, we emphasize that the calculation in this
paper eliminates the largest existing source of theoretical uncertainty
by taking into account leading logarithms of $p_T^2/m_c^2$ at all orders
in $\alpha_s$.

We have combined the LP corrections that we have calculated with the NLO
fixed-order calculations from
Refs.~\cite{Ma:2010jj,Ma:2010yw,Chao:2012iv} to obtain predictions for
the production cross sections and polarizations as functions of $p_T$.
By fitting the cross-section predictions to the Tevatron and LHC
cross-section data, we have obtained values for the NRQCD
nonperturbative long-distance matrix elements (LDMEs) that enter into
the production predictions through order $v^4$. Since the LP
approximation is valid only for $p_T\gg m_H$, where $m_H$ is the
quarkonium mass, we use only data for which $p_T$ is greater than $3
m_H$. We obtain good fits to the high-$p_T$ cross sections, with
$\chi^2/{\rm d.o.f.}\ll 1$ in each case.

One interesting result of the fits to the $\chi_{cJ}$ cross sections is
that the value of the ${}^3P_0^{[1]}$ LDME that we obtain is in good
agreement with the value that has been obtained in a potential model and
with values that have been extracted from the two-photon decays of the
$\chi_{c0}$ and $\chi_{c2}$. This agreement of values of the
${}^3P_0^{[1]}$ LDME that have been obtained through very different
methods is important evidence in support of the NRQCD factorization
conjecture. We note that, in previous works on the $\chi_{cJ}$ cross
section, which were based on fixed-order NLO calculations, the
${}^3P_0^{[1]}$ LDME was fixed to values that were obtained from potential
models~\cite{Ma:2010vd, Li:2011yc,Gong:2012ug,Shao:2014fca,Jia:2014jfa,
Shao:2014yta}.

We have used our cross-section predictions to predict the ratio
$R_{\chi_c}$, which is the $\chi_{cJ}$ feeddown contribution to the
prompt-$J/\psi$ cross section divided by the prompt-$J/\psi$ cross
section itself. The prediction lies systematically below the data for
$p_T<15$~GeV. This discrepancy in $R_{\chi_c}$ seems to be the result of
a downward deviation in the numerator combined with an upward deviation
in the denominator. However, the predictions for both numerator and the
denominator agree with the data within uncertainties.

We have also used the extracted LDMEs to predict the $J/\psi$,
$\psi(2S)$, and $\chi_{cJ}$ polarizations. The predictions for the
$J/\psi$ polarizations agree with the CMS data and the CDF Run~I data,
but lie systematically above the CDF Run~II data. The CDF Run~I data
show a slightly longitudinal polarization, while the CMS data show a
slightly transverse polarization. However, the theoretical predictions
are very similar for the CDF and CMS kinematics. The predictions for the
$\psi(2S)$ polarizations agree with the Tevatron data and the LHC data,
although the theoretical and experimental uncertainties are quite large.
For $\psi(2S)$ production, the ${}^1S_0^{[8]}$ channel is no longer
dominant at large $p_T$, and, so, the predicted $\psi(2S)$ polarization
becomes more transverse as $p_T$ increases. 
It is important to test
this prediction through measurements of the $\psi(2S)$ polarization with
good precision at larger values of $p_T$. There are, as yet, no
measurements of the $\chi_{cJ}$ polarizations. These would also provide
very useful tests of the theoretical predictions.

While we have obtained a reasonably good description of the
hadroproduction cross sections and polarizations for the $J/\psi$,
$\psi(2S)$, and $\chi_{cJ}$ states, our results do not address two
outstanding problems in quarkonium production, namely, the HERA $J/\psi$
photoproduction cross section, as measured by the H1 Collaboration
\cite{Adloff:2002ex,Aaron:2010gz}, and the $\eta_c$ hadroproduction cross
section, as measured by the LHCb Collaboration \cite{Aaij:2014bga}. In
the case of the $J/\psi$ photoproduction cross section, additional LP
fragmentation corrections, analogous to those that were computed in this
paper, were computed in Ref.~\cite{Bodwin:2015yma}. Those additional LP
corrections have very small effects on the photoproduction cross
section. For the choices of LDMEs that were used in
Ref.~\cite{Bodwin:2015yma}, the theoretical prediction for the
photoproduction cross section is dominated by the contribution from the
${}^1S_0^{[8]}$ channel. The value for $\langle {\cal O}^{J/\psi}
({}^1S_0^{[8]}) \rangle$ in Eq.~(\ref{Jpsi-LDMEs}) is about 10\% larger
than the value from Ref.~\cite{Bodwin:2014gia}, which was used in
Ref.~\cite{Bodwin:2015yma}. Hence, it makes the discrepancy between
theory and experiment slightly worse. In the case of the $\eta_c$
cross section, the change in value of $\langle {\cal O}^{J/\psi}
({}^1S_0^{[8]}) \rangle$ from Ref.~\cite{Bodwin:2014gia} to the present
paper also makes the discrepancy between theory and experiment
slightly worse.

While there remain important discrepancies between theory and experiment
in quarkonium production at high $p_T$, the theoretical predictions are
far from settled. At a minimum, a complete calculation of all of the LP
contributions in order $\alpha_s^5$ is needed in order to have
reasonable control of the theoretical uncertainties. These LP
contributions in order $\alpha_s^5$ may be most important in the
${}^3P_J^{[8]}$ channel because of their greater potential to affect the
shape in that channel. Higher-order calculations of NLP contributions
may also be needed, especially in the ${}^1S_0^{[8]}$ channel, for which
the LP contributions are not dominant until very large values of $p_T$.
New measurements at the LHC of the $\psi(nS)$, $\chi_{cJ}$,
$\Upsilon(nS)$, and $\chi_{bJ}$ cross sections and polarizations and the
$\eta_c$ cross section, all at unprecedentedly large values of $p_T$, can
provide definitive tests of the improved theoretical predictions.

\begin{acknowledgments}

We thank Jean-Philippe Guillet for providing information about the
computer code that implements the NLO parton-scattering results of
Ref.~\cite{Aversa:1988vb}. The work of G.T.B.\ and H.S.C.\  
is supported by the
U.S.\ Department of Energy, Division of High Energy Physics, under
Contract No.\ DE-AC02-06CH11357. The work of K.-T.C.\ and Y.-Q.M.\ 
is supported in part
by the National Natural Science Foundation of China (Grants No.\ 11475005
and No.\ 11075002). 
J.L.\ thanks the Korean Future Collider Working Group for enjoyable
discussions regarding the work presented here.
The work of J.L.\ and U-R.K.\ was supported by the Do-Yak project of
National Research Foundation of Korea funded by the Korea government (MSIP)
under Contract No.\ NRF-2015R1A2A1A15054533.
The submitted manuscript has been created in part by
UChicago Argonne, LLC, Operator of Argonne National Laboratory. Argonne,
a U.S.\ Department of Energy Office of Science laboratory, is operated
under Contract No.\ DE-AC02-06CH11357.

\end{acknowledgments}
\appendix*
\section{Numerical treatment of divergences in fragmentation functions
\label{sec:appendix}
}

The LP-factorization contribution to the cross section is given by the
convolution of the PPCSs $d \hat \sigma_{AB\to i+X} / d p_T$ and the
FFs $D_{i \to Q \bar Q(n)} (z,\mu_f)$:
\begin{equation}
\frac{d \sigma^{\rm LP}_{AB \to Q \bar Q(n)+X}}{d p_T}
= \int_{z_0}^1 dz
\frac{d \hat \sigma_{AB \to i+X}(z,\mu_f)}{d p_T}
D_{i \to Q \bar Q(n)} (z,\mu_f).
\end{equation}
Here, $z_0 = \frac{p_T}{\sqrt{s}} (e^{+y} + e^{-y})$.
We compute the evolved FFs by solving the LO DGLAP equation in Mellin
space (moment space) and performing the inverse Mellin transform
numerically. As is discussed in Sec.~\ref{sec:DGLAP}, the inverse Mellin
transform becomes numerically unstable near $z=1$ because
$D_{i \to Q \bar Q(n)} (z,\mu_f)$ can vary rapidly in this region
and may even diverge at $z=1$.

In order to deal with this problem, we partition the integral over $z$ as
follows:
\begin{eqnarray}
\int_{z_0}^1 dz
\frac{d \hat \sigma_{AB \to i+X}(z,\mu_f)}{d p_T}
D_{i \to Q \bar Q(n)} (z,\mu_f)
&=&
\int_{z_0}^{1-\epsilon} dz
\frac{d \hat \sigma_{AB \to i+X}(z,\mu_f)}{d p_T}
D_{i \to Q \bar Q(n)} (z,\mu_f)
\nonumber \\ &&
+
\int_{1-\epsilon}^1 dz
\frac{d \hat \sigma_{AB \to i+X}(z,\mu_f)}{d p_T}
D_{i \to Q \bar Q(n)} (z,\mu_f),
\nonumber \\
\end{eqnarray}
where $\epsilon$ is a small, positive number that is chosen so
that the evolved FF $D_{i \to Q \bar Q(n)} (z,\mu_f)$ can be
computed reliably for $z < 1-\epsilon$. In order to compute the integral
over $1-\epsilon < z <1$, we use the fact that the PPCSs behave as $d
\hat \sigma_{AB \to i+X}/ d p_T \sim z^N$ for $z \approx 1$, where $N
\approx 4$. Hence, we have
\begin{eqnarray}
&& \hspace{-5ex}
\int_{1-\epsilon}^1 dz
\frac{d \hat \sigma_{AB \to i+X}(z,\mu_f)}{d p_T}
D_{i \to Q \bar Q(n)} (z,\mu_f)
\nonumber \\
&=&
\int_{1-\epsilon}^1 dz
\bigg[ \frac{d \hat \sigma_{AB \to i+X}(z,\mu_f)}{d p_T} z^{-N} \bigg]
[ z^N D_{i \to Q \bar Q(n)} (z,\mu_f) ]
\nonumber \\
&\approx&
\bigg[ \frac{d \hat \sigma_{AB \to i+X}(z,\mu_f)}{d p_T} \bigg]_{z=1}
\times
\int_{1-\epsilon}^1 dz
\, z^N D_{i \to Q \bar Q(n)} (z,\mu_f)
\nonumber \\
&=&
\bigg[ \frac{d \hat \sigma_{AB \to i+X}(z,\mu_f)}{d p_T} \bigg]_{z=1}
\times
\bigg[
\int_0^1 dz
\, z^N D_{i \to Q \bar Q(n)} (z,\mu_f)
-
\int_0^{1-\epsilon} dz
\, z^N D_{i \to Q \bar Q(n)} (z,\mu_f)
\bigg]
\nonumber \\
&=&
\bigg[ \frac{d \hat \sigma_{AB \to i+X}(z,\mu_f)}{d p_T} \bigg]_{z=1}
\times
\bigg[
\tilde D_{i \to Q \bar Q(n)} (N+1,\mu_f)
-
\int_0^{1-\epsilon} dz
\, z^N D_{i \to Q \bar Q(n)} (z,\mu_f)
\bigg],
\end{eqnarray}
where we have expanded $z^{-N} d \hat \sigma_{AB \to i+X}/d p_T$
in powers of $1-z$ and retained only the leading-order contribution,
which is simply the value of $d \hat \sigma_{AB \to i+X}/d p_T$ at
$z=1$. The quantity $\tilde D_{i \to Q \bar Q(n)} (N+1,\mu_f)$,
$(N+1)$st moment of $D_{i \to Q \bar Q(n)}$, is known analytically, and
the integral over the range $0 < z < 1-\epsilon$ can be computed
numerically. Hence, in our calculations, we use the expression
\begin{eqnarray}
&& \hspace{-5ex}
\int_{z_0}^1 dz
\frac{d \hat \sigma_{AB \to i+X}(z,\mu_f)}{d p_T}
D_{i \to Q \bar Q(n)} (z,\mu_f)\nonumber
\\
&\approx&
\int_{z_0}^{1-\epsilon} dz
\frac{d \hat \sigma_{AB \to i+X}(z,\mu_f)}{d p_T}
D_{i \to Q \bar Q(n)} (z,\mu_f)
\nonumber \\ &&
+
\bigg[ \frac{d \hat \sigma_{AB \to i+X}(z,\mu_f)}{d p_T} \bigg]_{z=1}
\times
\bigg[
\tilde D_{i \to Q \bar Q(n)} (N+1,\mu_f)
-
\int_0^{1-\epsilon} dz
\, z^N D_{i \to Q \bar Q(n)} (z,\mu_f)
\bigg]. \nonumber \\
\label{evolution-espression}
\end{eqnarray}
In our numerical calculations, we take $N=4$ and $\epsilon=10^{-6}$. We
have varied $N$ between $3.1$ and $7$ and find that the largest
sensitivity to $N$ occurs at low $p_T$ and is less than $3\times
10^{-6}$ of the contribution in each channel.

We have compared numerical results from Eq.~(\ref{evolution-espression})
for $\mu_f$ near $\mu_0$ with the analytic expression for the evolved
FFs through second order in $\alpha_s$. The results agree to better than
$1\%$. We expect numerical difficulties in
Eq.~(\ref{evolution-espression}) to be most severe as $\mu_f$ approaches
$\mu_0$, where the evolved FFs approximate the initial FFs, which, in
some cases, are distributions at $z=1$. Hence, good agreement with the
analytic expressions in this region gives us confidence that the
algorithm that is based on Eq.~(\ref{evolution-espression}) is reliable.


\end{document}